\documentclass[preprint,superscriptaddress,nofootinbib]{revtex4}

\usepackage{graphicx, hyperref, amsmath, amssymb, slashed, color, bbm,enumitem}
\usepackage{color}

\newcommand{\nc}{\newcommand}

\nc{\beq}{\begin{equation}}
\nc{\eeq}{\end{equation}}
\nc{\barray}{\begin{eqnarray}}
\nc{\earray}{\end{eqnarray}}
\nc{\barrayn}{\begin{eqnarray*}}
\nc{\earrayn}{\end{eqnarray*}}
\newcommand{\bea}{\begin{eqnarray}}
\newcommand{\eea}{\end{eqnarray}}
\nc{\bcenter}{\begin{center}}
\nc{\ecenter}{\end{center}}
\nc{\ket}[1]{| #1 \rangle}
\nc{\bra}[1]{\langle #1 |}
\nc{\0}{\ket{0}}
\nc{\mc}{\mathcal}
\nc{\er}[1]{(\ref{eq:#1})}
\nc{\onehalf}{\frac{1}{2}}
\nc{\partialbar}{\bar{\partial}}
\nc{\psit}{\widetilde{\psi}}
\nc{\Tr}{\mbox{Tr}}
\nc{\hc}{\mbox{H.c.}}
\nc{\ev}{\;\mathrm{eV}}
\nc{\mev}{\;\mathrm{MeV}}
\nc{\gev}{\;\mathrm{GeV}}
\nc{\tev}{\;\mathrm{TeV}}

\def\chii0{\chi_i^0}
\def\chij0{\chi_j^0}

\newcommand{\gsim}{\lower.7ex\hbox{$\;\stackrel{\textstyle>}{\sim}\;$}}
\newcommand{\lsim}{\lower.7ex\hbox{$\;\stackrel{\textstyle<}{\sim}\;$}}
\nc{\ttbar}{t\bar t}
\nc{\TAFB}{A_{FB}^t}
\nc{\lepAFB}{A_{FB}^\ell}
\nc{\TAC}{A_{C}^t}
\nc{\Lag}{\mathcal{L}}
\nc{\Proj}{\mathcal{P}}

\newcommand{\Eq}[1]{Eq.~(\ref{#1})}

\newcommand{\Fig}[1]{Fig.~\ref{#1}}

\begin{document}

\title{NMSSM Interpretation of the Galactic Center Excess}

\preprint{CALT 2014-157, MCTP-14-22}

\author{Clifford Cheung}
\affiliation{Walter Burke Institute for Theoretical Physics, California Institute of Technology, Pasadena, CA 91125, USA}

\author{Michele Papucci}
\affiliation{Michigan Center for Theoretical Physics, University of Michigan, Ann Arbor, MI 48109, USA}

\author{David Sanford}
\affiliation{Walter Burke Institute for Theoretical Physics, California Institute of Technology, Pasadena, CA 91125, USA}

\author{Nausheen R. Shah}
\affiliation{Michigan Center for Theoretical Physics, University of Michigan, Ann Arbor, MI 48109, USA}

\author{Kathryn M. Zurek}
\affiliation{Michigan Center for Theoretical Physics, University of Michigan, Ann Arbor, MI 48109, USA}


\begin{abstract}
We explore models for the GeV Galactic Center Excess (GCE) observed by the Fermi Telescope, focusing on $\chi \chi \rightarrow f \bar f$ annihilation processes in the $Z_3$ NMSSM.  We begin by examining the requirements for a simplified model (parametrized by the couplings and masses of dark matter (DM) and mediator particles) to reproduce the GCE via $\chi \chi \rightarrow f \bar f$, while simultaneously thermally producing the observed relic abundance.  We apply the results of our simplified model to the $Z_3$ NMSSM for Singlino/Higgsino~(S/H) or Bino/Higgsino~(B/H) DM. In the case of S/H DM, we find that the DM must be be very close to a pseudoscalar resonance to be viable, and large $\tan\beta$ and positive values of $\mu$ are preferred for evading direct detection constraints while simultaneously obtaining the observed Higgs mass. In the case of B/H DM, by contrast, the situation is much less tuned: annihilation generally occurs off-resonance, and for large $\tan\beta$ direct detection constraints are easily satisfied by choosing $\mu$ sufficiently large and negative.  The B/H model generally has a light, largely MSSM-like pseudoscalar with no accompanying charged Higgs, which could be searched for at the LHC. 
\end{abstract}

\maketitle

\section {Introduction}

In recent years, an intriguing excess of $\sim$ 1--3 GeV gamma ray
photons has appeared in the galactic center
\cite{Goodenough:2009gk,Hooper:2010mq,Abazajian:2012pn,Gordon:2013vta}. This
galactic center excess (GCE) is approximately spherically symmetric,
with a spatial distribution consistent with annihilating dark
matter~(DM) following an Navarro-Frenk-White (NFW) profile
\cite{Hooper:2013nhl,Daylan:2014rsa}. As is often true of signals from
indirect detection, it is not clear whether the GCE is a hint of
physics beyond the standard model (BSM) or is of astrophysical
origin~\cite{Petrovic:2014uda,Carlson:2014cwa}. Given astrophysical
uncertainties, it is worth exploring the DM hypothesis to assess how
difficult it is to build theories which can accommodate the excess.
Given concrete models, one can then make predictions that can be
tested in more controlled environments such as particle colliders and
DM direct detection experiments.

The GCE is well fit by a $\sim$ 30--40 GeV DM particle annihilating
directly into $b\bar b$ with a cross-section of order $\langle \sigma
v \rangle \simeq 2 \times 10^{-26} \textrm{ cm}^3/\textrm{s}$, which
is intriguingly close to that of a thermal relic.  Annihilation to
$\tau\bar \tau$ can also fit the data, though not as well and for a
lower DM mass of $\sim 10$ GeV.  Already, there has been much work
done to understand possible underlying particle physics models of this
DM interpretation
\cite{Abdullah:2014lla,Basak:2014sza,Berlin:2014pya,Boehm:2014bia,Cline:2014dwa,Ghosh:2014pwa,Ipek:2014gua,Ko:2014gha,Martin:2014sxa,Wang:2014elb,Arina:2014yna,Izaguirre:2014vva,Modak:2013jya,Detmold:2014qqa}.
While there is some tension between the GCE and constraints from
anti-proton bounds on dark matter
annihilation~\cite{Kong:2014haa,Bringmann:2014lpa}, the GCE region
remains allowed for the $b\bar b$ channel for conservative choices of
propagation model.

Simplified models of DM  describing the particles and interactions undergoing annihilation processes via $\chi\chi\rightarrow b \bar b$ are a useful tool for obtaining a handle on the underlying dynamics of the interaction.  
Such a process may be mediated by (colored) $t$-channel or (neutral) $s$-channel particles.  The former are strongly constrained by LEP and LHC data.  As a consequence, we focus throughout on $s$-channel mediators.  As noted in \cite{Huang:2013apa,Boehm:2014hva,Ipek:2014gua}, pseudoscalar $s$-channel mediators are well-suited because they are not immediately excluded by direct detection experiments.  Using this simplified model, we can determine the coupling strengths and masses required to fit all of the experimental data, including a careful analysis of the relic abundance in such a theory.

With this simplified model analysis in hand, one can apply the needed features of the theory to particular models of DM.  Supersymmetric extensions of the standard model (SM) are
a well-motivated class of renormalizable models which can accommodate
a stable DM particle together with new degrees of freedom to mediate
annihilation.  Unfortunately, it is not possible to explain the GCE
within the minimal supersymmetric standard model (MSSM) via $s$-channel annihilation through a pseudoscalar\footnote{Annihilation through $t$-channel scalars in the MSSM is also strongly constrained, as we describe in Sec.~\ref{Sec:Simplified} below.}.  The reason
is simple: the GCE requires light DM, but a thermal relic
abundance demands an MSSM pseudoscalar that is too light to be
consistent with existing LHC constraints \cite{ATLAS-CONF-2013-090,Chatrchyan:2012vca}. 
These constraints are  derived from charged Higgs searches and precision Higgs constraints.  Progress can thus be made by decoupling the pseudoscalar mass from the charged and CP-even heavy Higgs masses.
The simplest MSSM extension satisfying this requirement is the next-to-minimal
supersymmetric standard model (NMSSM). 
The NMSSM is a theoretically
well-motivated framework that offers all the necessary elements for
neutralino DM annihilating via $\chi \chi \rightarrow b \bar b$.  

The purpose of this paper is to show that the NMSSM can indeed generate the GCE via $2 \rightarrow 2$ annihilation while
evading stringent constraints on Higgs phenomenology from the Large
Hadron Collider (LHC) and null results from the LUX direct detection
experiment~\cite{Akerib:2013tjd}.  Constructing a working theory---that is, one with
thermal relic DM accommodating the GCE and consistent with existing
bounds---entails model building challenges which have not been
sufficiently emphasized in earlier works~\cite{Han:2014nba,Berlin:2014pya,Cerdeno:2014cda}.
To summarize, the primary results of this paper are as follows:
\begin{itemize}[leftmargin=.5cm,rightmargin=.5cm] 

\item An analysis of the simplified model for $\chi\chi \rightarrow b
  \bar b$ shows that resonant annihilation can significantly
  complicate models for the GCE.  In particular, theories with
  resonant annihilation predict a large discrepancy between the
  annihilation rate today as compared to the early universe.  Since
  the observed GCE annihilation cross-section is near that of a
  thermal relic, resonant models generically have difficulty explaining the
  GCE while maintaining a thermal relic.  As we will show, this
  difficulty can be overcome if there is a large hierarchy between the
  couplings of the mediating particle to final state fermions and the
  DM.  Alternatively, the presence of additional $\chi\chi \rightarrow
  b \bar b$ annihilation channels, particularly via the $Z$ boson, can
  alleviate the tension.

\item An analysis of the NMSSM reveals several surmountable model
  building challenges for explaining the GCE.  There are three main
  issues.  First, a complete model will often contain a scalar
  partner to the pseudoscalar that will mediate dangerous spin
  independent (SI) DM-nucleon scattering.  However stringent direct
  detection constraints can be alleviated if this new scalar is
  sufficiently heavy~\cite{Ipek:2014gua}, or if there is destructive
  interference---{\it a.k.a.}~blind spots---induced between different
  SI scattering processes.  Second, many of these models induce mixing
  between new scalars and the SM Higgs boson, modifying SM Higgs decay
  modes in a way that may be in conflict with LHC constraints\footnote{See Ref.~\cite{Curtin:2013fra} for a detailed study of possible exotic decays of the 125 GeV Higgs.}.  Third,
  if any component of DM carries electroweak charges, then
  $Z$-mediated $p$-wave suppressed annihilation in the early universe
  can be quite important, thus offsetting the correlation between the
  GCE and the thermal relic cross-section, which may be problematic in
  models where the abundance is set via non-resonant annihilation.  

\item We have identified a parameter space of the $Z_{3}$ NMSSM which can
  accommodate the GCE while simultaneously evading all collider and
  direct detection constraints.  These models are 1) Singlino/Higgsino
  DM via resonant annihilation through the pseudoscalar, or 2)
  Bino/Higgsino via off-resonant annihilation through the
  pseudoscalar.  In both cases, most of the parameter space is
  accessible at the next generation of direct detection experiments. The latter case
  also provides interesting phenomenological consequences for the LHC Run II deserving
  further investigation.

\end{itemize}

For this paper we have used semi-analytical methods to study the relevant parameter space. All couplings and cross-section were output using {\tt CalcHEP 3.4}~\cite{Belyaev:2012qa}. We checked our analytic results thoroughly using {\tt micrOMEGAs}~\cite{Belanger:2013oya} and {\tt NMSSMTools}~\cite{Ellwanger:2004xm, Ellwanger:2005dv, Belanger:2005kh, Das:2011dg, Muhlleitner:2003vg} where applicable. Our paper is organized as follows. In Sec.~\ref{Sec:Simplified} we summarize a simplified model for DM annihilation via a pseudoscalar, enumerating the conditions needed to accommodate a thermal relic density simultaneously with the GCE. In Sec.~\ref{Sec:NMSSM} we present an analysis of the $Z_3$ NMSSM, detailing characteristics of the neutralino DM and the required properties of the scalar and pseudoscalar sectors to give a cosmologically viable model. We reserve Sec.~\ref{Sec:Conc} for our conclusions. The full detailed analytic formulae pertaining to both the general and the $Z_3$ NMSSM are presented in the appendices.

\section{Simplified Model Analysis}
\label{Sec:Simplified}

In this section we present a simplified model for a thermal relic DM
candidate consistent with the GCE.  Throughout, we assume Majorana DM
that annihilates through the hadronic channel, $\chi \chi \rightarrow
b \bar b$,with a DM mass in the range $\sim$ 30--40 GeV, as preferred
by the fits in Ref.~\cite{Daylan:2014rsa}.   One can also
consider leptonic annihilation via $\chi\chi \rightarrow \tau \bar
\tau$, though the fit for this channel is poorer; we will not
consider it further.

A priori, $\chi \chi \rightarrow b \bar b$ scattering can be mediated
via $s$-channel or $t$-channel exchange.  If the mediator is
in the $t$-channel, then it must be colored.  To accommodate a thermal
relic abundance, the mediator must be quite light, with mass $\sim
100~\gev$, which is in tension with stringent LEP and LHC limits on colored
particles decaying to DM particles and $b$-jets, unless the mediator and the DM particle are very degenerate in mass.  
For example, neutralino annihilation via $t$-channel light ($\lesssim 100$ GeV) sbottom exchange is
highly constrained in the MSSM.  Even if sbottom mixing angles can be
tuned to evade stringent LEP constraints~\cite{Batell:2013psa}, direct limits 
on colored production of the heavier sbottom are strong, and not obviously surmountable. 
Furthermore, in the sbottom-neutralino degenerate case, co-annihilation 
in the early universe play an important role in setting the relic abundance, 
requiring different neutralino annihilation cross sections than those preferred by the GCE.

Consequently, we restrict ourselves to an $s$-channel mediator which is a
vector, scalar, or pseudoscalar.  In all cases we consider the case where DM is a Majorana fermion, resulting in a factor of 4 difference in relevant cross-sections as compared to a Dirac fermion.  If the mediator
is a gauge boson or a scalar, then DM annihilation is $p$-wave suppressed and thus
negligible in the present day.   Thus,
we focus on the case where the mediator is a pseudoscalar, which we
denote by $a$, and which was considered in Refs.~\cite{Huang:2013apa,Boehm:2014hva}.  

Considering only the coupling to
$b\bar{b}$ needed for the GCE, the simplified model describing the
coupling of a Majorana DM particle $\chi$ coupled to $a$ has the
interaction Lagrangian,
\begin{eqnarray}
-{\cal L}_{\rm int} &=& i y_{a \chi\chi} a \bar \chi \gamma^5 \chi +i
y_{a bb} a \bar b \gamma^5 b .
\label{eq:simpL}
\end{eqnarray}
Consequently, the entire parameter space of the model is fixed by the pseudoscalar and DM masses, $m_a$ and $m_\chi$, and the dimensionless Yukawa couplings, $y_{a \chi\chi}$ and $y_{a b b}$.  The present day DM annihilation cross-section is
\begin{eqnarray}
\sigma v \big|_{v=0} &\simeq& \frac{3}{2\pi} \frac{y_{a \chi\chi}^2 y_{a b b}^2 m_\chi^2 }{(m_a^2 - 4m_\chi^2)^2 + m_a^2 \Gamma_a^2 }\, ,
\label{eq:simpsigma}
\end{eqnarray}
where $\Gamma_a$ is the decay width of the pseudoscalar mediator $a$,
\begin{eqnarray}
\Gamma_a &\simeq & \frac{ m_a}{16\pi} (y_{a \chi\chi}^2 +6y_{a bb}^2).
\label{eq:Gamma}
\end{eqnarray}
The pseduoscalar $a$ may have decay modes to other light SM fermions, but
these must be chirality suppressed to satisfy flavor bounds, so we
neglect them. As noted in~\cite{Daylan:2014rsa}, the DM annihilation cross-section
inferred from the GCE is of order $ \sigma v \big|_{v=0} \simeq 2
\times 10^{-26} \textrm{ cm}^3/\textrm{s}$, which is, remarkably,
within the ball park of a thermal relic cross-section.  Additionally,
because $a$ is a pseudoscalar, it cannot mediate spin-independent
DM-nucleon scattering, and thus this simplified model automatically
avoids direct detection bounds.

Given the observed GCE annihilation cross-section, it is tempting to
assume that the same annihilation process also mediated thermal
freeze-out in the early universe.  Such a setup works well
in the case that the annihilation is not resonant, {\it i.e.}~when
$m_a$ and $2m_\chi$ are not highly degenerate.  To test this
condition it is useful to define a degeneracy parameter,
\begin{eqnarray}
\delta &=& |1- 4 m_\chi^2 / m_a^2|,
\end{eqnarray} 
which characterizes the proximity of the theory to the resonant regime.  If $\delta$ is not very small, then the annihilation is not resonant,
and the GCE and a thermal relic abundance can be simultaneously accommodated as long as the product $y_{a \chi\chi}^2 y_{a bb}^2$ is fixed to an appropriate value:
\begin{equation}
\sigma v \simeq 2 \times 10^{-26} \mbox{ cm}^3/\mbox{s} \left(\frac{y_{a b b}}{y_b}\right)^2 \left(\frac{y_{a \chi \chi}}{0.6}\right)^2 \left(\frac{m_\chi}{35 \mbox{ GeV}}\right)^2\left(\frac{(120 \mbox{ GeV})^2-4(35\mbox{ GeV})^2}{m_a^2-4 m_\chi^2}\right)^2,
\label{OffResonance}
\end{equation}
where $y_b$ is the SM bottom quark Yukawa.

However, the story changes substantially if $\delta \sim 0$, in which
case annihilation is resonant.  As is well-known~\cite{Griest:1990kh},
resonant DM annihilation will be substantially different today as compared to
the early universe.  This happens because of thermal broadening of the resonance during the process of DM freeze-out.  From~\cite{Griest:1990kh}, the resonant annihilation cross-section at a given $x = m_{\chi}/T$ is  
\begin{eqnarray}
\langle \sigma v \rangle 
&\simeq & \frac{3 e^{-x \delta}    x^{3/2}\delta^{1/2}  y_{a \chi\chi}^2 y_{a bb}^2 m_\chi^2 }{\sqrt{\pi} m_{a}^3 \Gamma_a }.
\label{eq:sigma}
\end{eqnarray}
Integrating over $x$ gives the relic abundance
\begin{equation}
\Omega h^2 = \frac{3.12 \times 10^{-12}  m_a^3 \Gamma_a}{ (\textrm{GeV})^2m_{\chi}^2 y_{a \chi\chi}^2 y_{a bb}^2 \mbox{Erfc}\left[\sqrt{x_f \delta}\right]},
\label{eq:Omega}
\end{equation} 
where $x_f$ is the value of $x$ at freeze-out.  Plugging the decay width in \Eq{eq:Gamma} into the GCE cross-section, in the limit when the width is controlling the cross-section, for $2 m_\chi < m_a$
we find that 
\begin{eqnarray}
\sigma v \big|_{v=0} &\sim& 2\times10^{-26}\text{cm}^3 \left(\frac{4
  m_{\chi }^2}{m_a^2}\right)\left(\frac{70 \text{
    GeV}}{m_a}\right)^2\left(\frac{10^{-3}} { \frac{y_{a\chi \chi } }{
    y_{abb}}\frac{\delta}{6}+\frac{y_{abb}}{y_{a\chi \chi
}}}\right)^2.
\end{eqnarray}
Similarly, the relic abundance close to resonance can be written as:
\begin{eqnarray}
\Omega h^2 &\sim&0.12 \left(\frac{m_a^2}{4
  m_\chi^2}\right)\left(\frac{m_a}{70\text{ GeV}}\right)^2\left[\frac{
    y_{a\chi\chi}^{-2}+(\delta/ 6) \,y_{abb}^{-2}}{10^{6}}\right] \left(\frac{\textrm{Erfc}[1.325]}{\textrm{Erfc}\left[\sqrt{x_f}\,\delta\right]}\right).
\end{eqnarray}
Thus the relic abundance is controlled by the smaller of $y_{a bb}$
and $y_{a \chi\chi} \sqrt{\delta / 6}$.  On the other hand, present
day DM annihilation is controlled by the larger of $y_{a bb} / y_{a
  \chi\chi}$ and $y_{a\chi\chi} \delta / 6 y_{a bb}$.

\begin{figure}[!p]
\begin{tabular}{cc}
\includegraphics[width=0.45\textwidth]{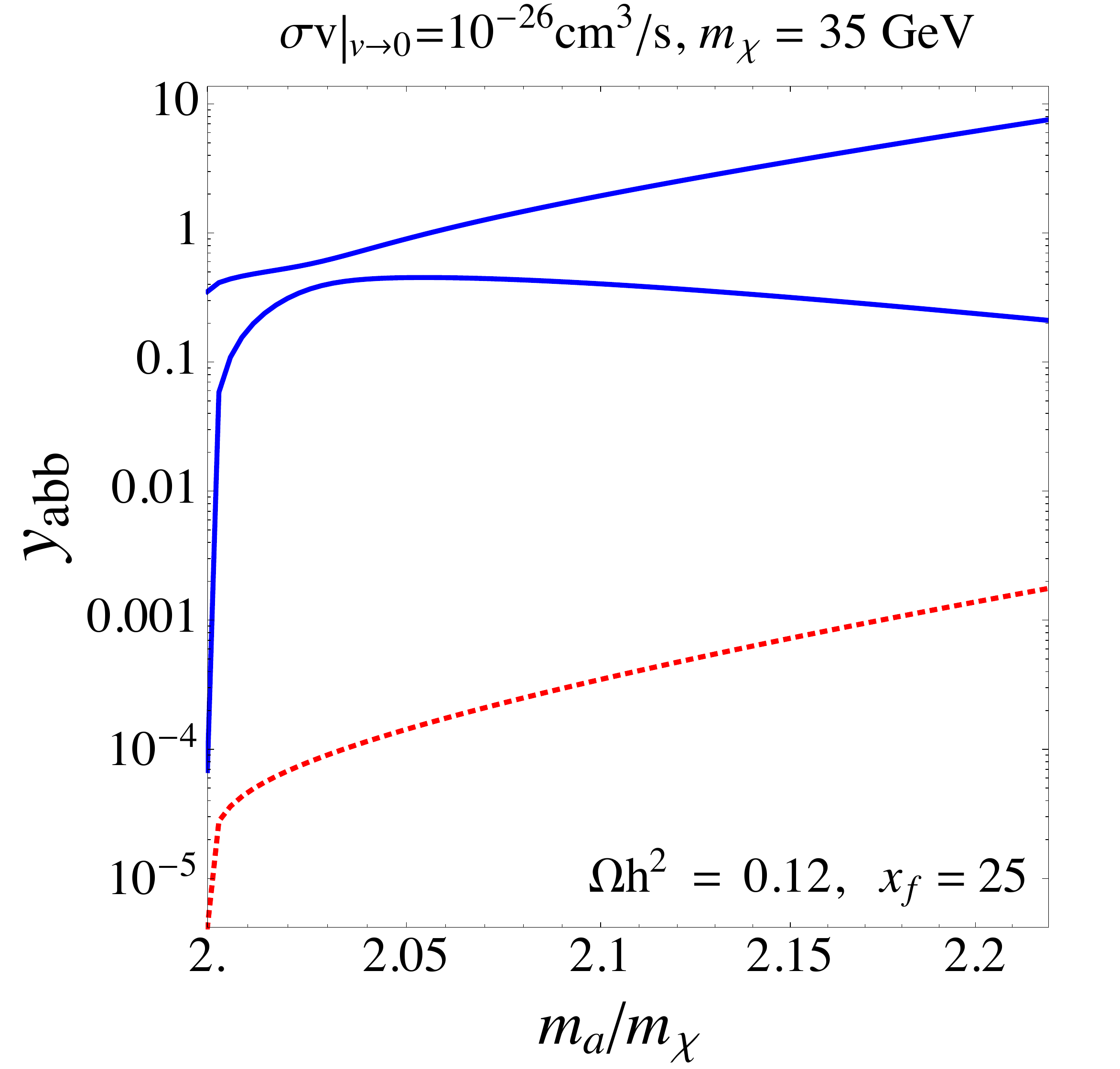} &
\includegraphics[width=0.45\textwidth]{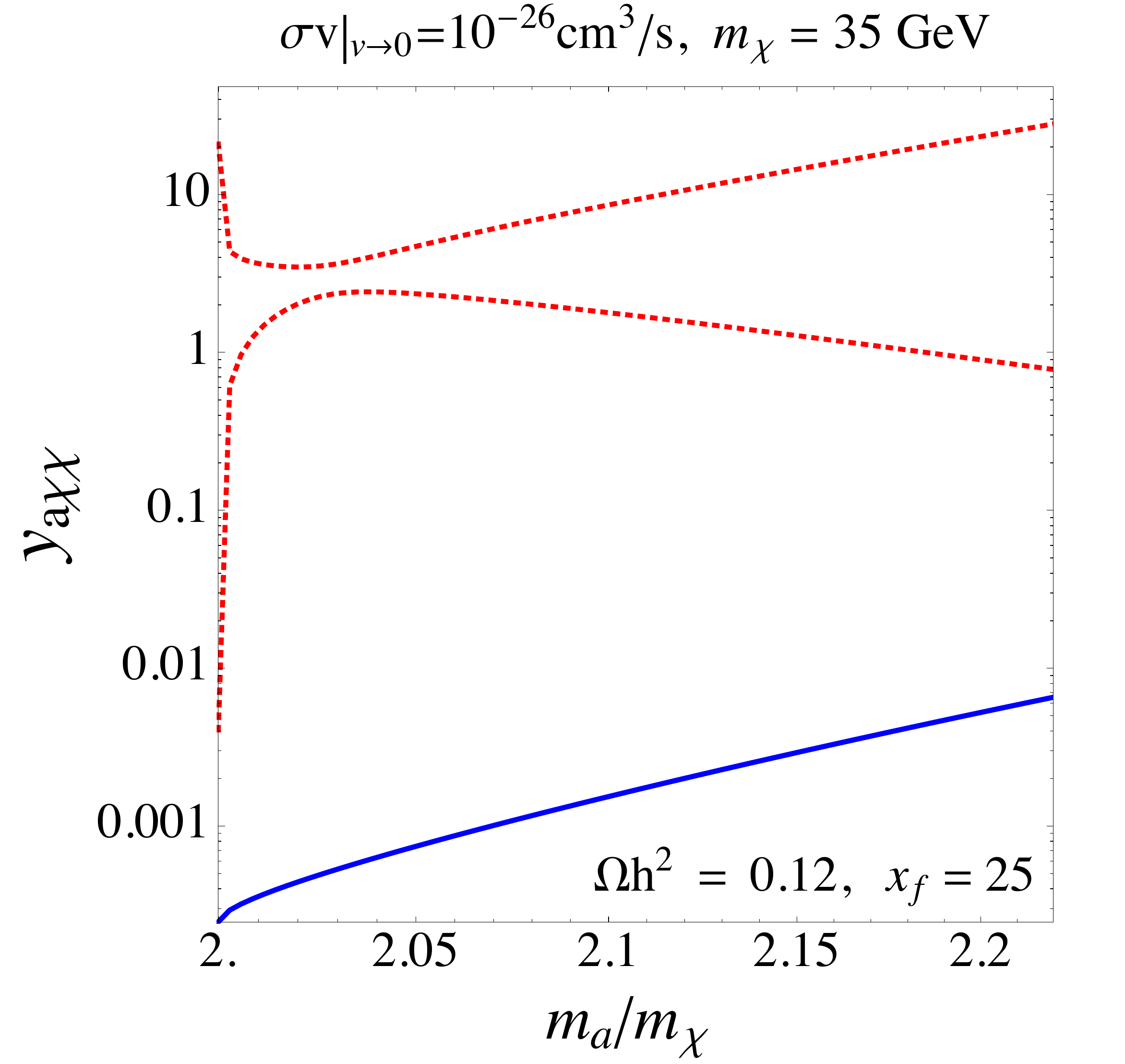} \\
\includegraphics[width=0.45\textwidth]{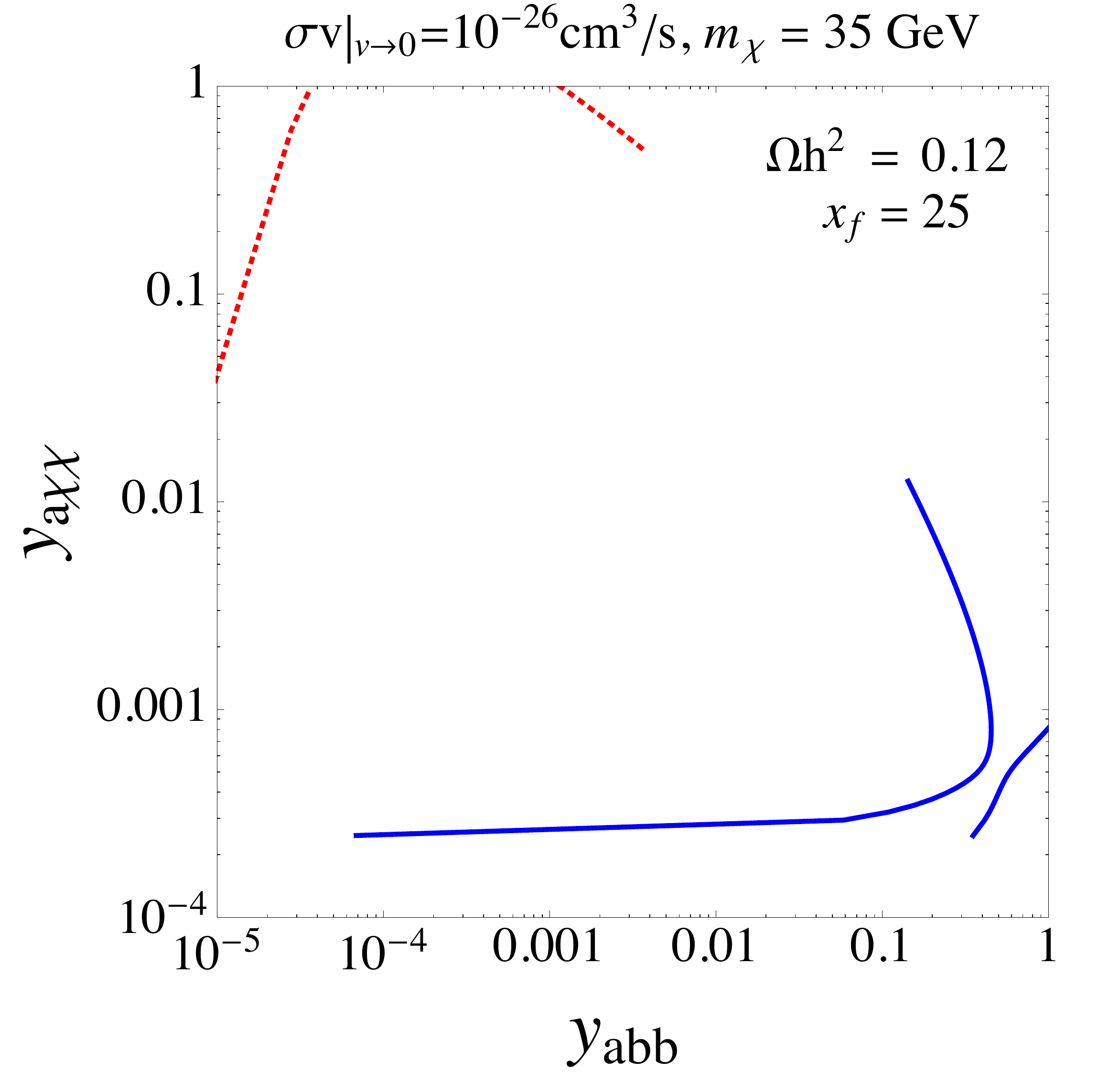} &
\includegraphics[width=0.45\textwidth]{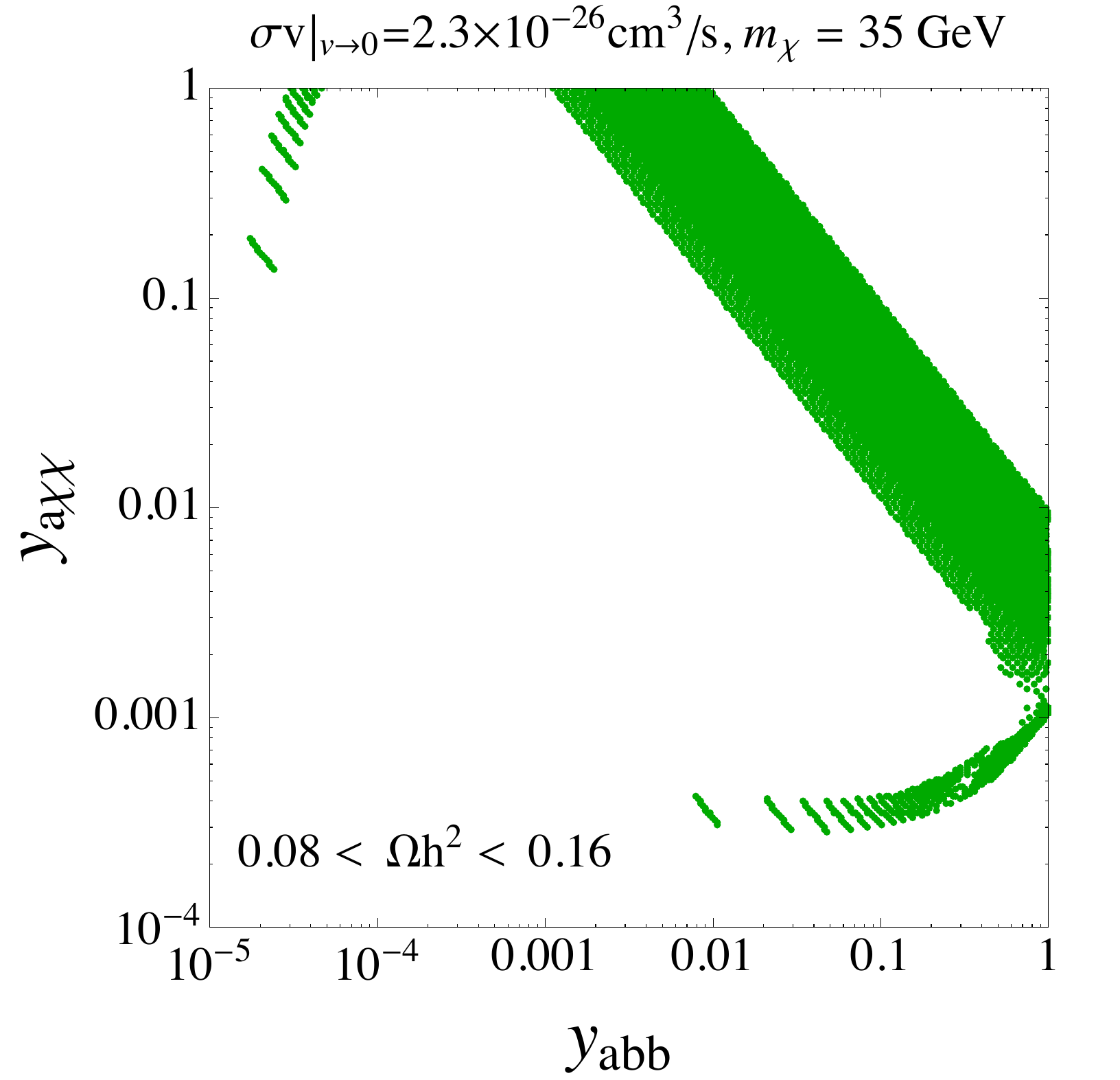}
\end{tabular}
\caption{{\em Upper panels}: Couplings $y_{a bb}$ and $y_{a \chi
    \chi}$ required to obtain relic density and GCE through resonant
  annihilation as a function of $m_a/m_\chi$.  The left and right
  panels should be read as a pair, with blue (solid) and red (dashed) curves in the
  left panel coupling to the similarly denoted curve in the right panel.  In
  the left panel, either blue (solid) curve can be matched with the single
  blue (solid) curve in the right panel, and likewise either red (dashed) curve in the
  right panel can be matched with the single red (dashed) curve in the left
  panel to obtain both the observed relic abundance and the GCE.  \\{\em
    Lower panels}: Allowing $m_a/m_\chi$ to float, the allowed
  couplings $y_{a bb}$ and $y_{a \chi \chi}$ to obtain both the relic
  abundance and the GCE are shown, obtained using analytic results
  (left panel) and the full numerical scan (right panel). The blue~(solid) and red~(dashed) curves in the left panel correspond to the similarly denoted curves in the top two panels.  One can see
  broad agreement between the analytic results and the output of the
  scan.  } \label{FreezeoutAndGC}
\end{figure}

\Fig{FreezeoutAndGC} depicts the couplings $y_{a \chi\chi}$ and $y_{a bb}$ required to simultaneously accommodate the observed DM relic abundance and GCE for a fixed DM mass of $m_\chi \sim 35$ GeV.  According to the upper two panels of \Fig{FreezeoutAndGC}, a thermal relic abundance requires that the smaller of $y_{a \chi\chi}$ and $y_{a bb}$  be of order $10^{-3}$.     Consequently, at least one of the couplings of the pseudoscalar mediator must be small.  On the other hand, the GCE annihilation cross-section of order $\sigma v|_{v=0} \sim 10^{-26}$ cm$^3$/s demands a ratio of order  $\sim 10^{3}$ between the two relevant couplings.  In other words, accommodating the GCE with resonant annihilation requires a large hierarchy between the couplings of $a$.   The story will change, however, if there are additional annihilation modes for the DM.  

More than $\sim 20\%$  away from resonance, \Eq{eq:sigma} and \Eq{eq:Omega} do not apply, but as expected,  $\sigma v\big|_{v\rightarrow 0}$ is correlated in the usual way with $\Omega h^2$.  To interpolate consistently between the resonant and non-resonant regimes, we have implemented this simplified model in {\tt micrOMEGAs\_3.6.7}~\cite{Belanger:2014hqa} to numerically scan over the couplings and mass of the scalar, fixing  $\sigma v|_{v\rightarrow 0} = 2.3 \times 10^{-26}$ cm$^3$/s, and DM mass to 35 GeV. The relic density obtained is shown in the lower two panels of Fig.~\ref{FreezeoutAndGC}, with the left panel being obtained analytically in the resonant regime and the right panel being the result of a numerical scan, which matches the analytic results.

\section{NMSSM Analysis}
\label{Sec:NMSSM}

We now apply the results of the simplified model in the previous section to the parameter space of the NMSSM.
In the appendices, we present our conventions and analytic
formulae, including the scalar and pseudoscalar masses and couplings
to the DM.  Throughout this analysis we restrict ourselves to the
$Z_3$ NMSSM, which has a superpotential
\begin{equation}
W = \lambda S H_u H_d  + \frac{1}{3}\kappa S^3,
\end{equation}
with soft breaking terms
\begin{equation}
-{\cal L}_{\rm soft} = \lambda A_\lambda S H_u H_d + \frac{1}{3}
\kappa A_\kappa S^3.
\end{equation}
The Peccei-Quinn symmetry limit is defined as $\kappa \rightarrow 0$.  There is of course more freedom in the general NMSSM, which gives greater
parameter freedom in the scalar sector to satisfy constraints, but the
$Z_3$ NMSSM is sufficient to study sample cases of viable regions.

As is well-known, for sufficiently large values of $\lambda$, the NMSSM Higgs mass can be substantially boosted from its usual mass range in the MSSM~\cite{Barbieri:2006bg,Hall:2011aa} without the need for very heavy stop squarks. However, this mass enhancement is only effective at small $t_\beta$, which we will find to be important later.

Within the NMSSM, there are three basic phenomenologically viable
neutralino identities: Singlino, Singlino/Higgsino (S/H), and
Bino/Higgsino (B/H).  The pure Singlino case is inaccessible in the
$Z_3$ NMSSM, as it requires vanishing $\lambda$ which implies $\mu =
0$.  For the $Z_3$-invariant NMSSM, a light, mostly
Singlino DM implies that $\kappa/\lambda \sim m_{\chi} /2 \mu $ must
be fairly small, since $\mu \gtrsim 150 \mbox{
  GeV}$ from LHC bounds (see Sect.~\ref{sec:SH}). 
  In addition, when Higgsino is mixed with Singlino,
annihilation through the $Z$ pole is opened, significantly modifying
both the relic density and current annihilation rate in the relevant
mass range to explain the GCE.  Annihilation through the $Z$-pole can
still be a factor even for points maintaining consistency with the LEP
constraints on the invisible width of the $Z$, bounding the Higgsino fraction to be small or $t_{\beta}$ to be close to 1 as discussed in detail later in Sect.~\ref{sec:SH}. Since the Higgsino fraction is set by $(\lambda
v_{u,d}/ \mu)$ this implies that $\lambda$ must also be kept fairly small.  In the
Bino/Higgsino case, by contrast, $\kappa/\lambda$ is taken large to
decouple the Singlino component. Since $\kappa$ is bounded by perturbativity constraints to be at most $\mathcal{O}(1)$, this forces $\lambda$ to be much smaller.

Given that $\kappa/\lambda \ll 1$ in the Singlino case, the greatest
challenge is to maintain a healthy CP-even sector.  This can be easily
understood upon diagonalizing the $H_{u},\,H_{d}$ sector to the $(H,h)$ (approximate) mass eigenstates  defined by $\langle h \rangle=v$, $\langle H \rangle=0$ (which correspond to the mass eigenstates in the MSSM decoupling limit), while keeping the singlet in the interaction basis\footnote{We will refer to this basis as the $(H,h,S)$ basis, in contrast to the $(H_{u},H_{d},S)$ interaction basis and the $(H,h,h_{S})$ mass basis.}. We identify $h$ with the SM-like
Higgs and $H$ with the heavier MSSM-like
Higgs.
In this basis the CP-even mass matrix is
\begin{eqnarray}
\mathcal{M}^2_h & = & \left(
\begin{array}{ccc}
m_A^2+s^2_{2\beta} \left(m_Z^2-\lambda ^2 v^2\right) & s_{2\beta}
c_{2\beta} \left(m_Z^2-\lambda ^2 v^2\right) & -\lambda v \mu
c_{2\beta} \left(\frac{m_A^2}{2 \mu^2 } s_{2\beta}+\frac{\kappa
}{\lambda }\right) \\
& c^2_{2\beta} m_Z^2+\lambda ^2 v^2 s^2_{2\beta} & 2\lambda v\mu
\left(1-\frac{m_A^2 }{4 \mu^2 }s^2_{2\beta}-\frac{\kappa }{2 \lambda }
s_{2\beta}\right) \\
& & \lambda ^2 v^2 s_{2\beta } \left(\frac{m_A^2 
  s_{2\beta }}{4\mu ^2}-\frac{\kappa }{2\lambda }\right)+\frac{\kappa \mu
  A_{\kappa }}{\lambda }+\frac{4 \kappa ^2 \mu ^2}{\lambda ^2}
\end{array}
 \right),\nonumber\\
\end{eqnarray}
where $t_\beta\equiv\tan\beta=v_u/v_d$, $s_\beta\equiv\sin\beta$, $c_\beta\equiv\cos\beta$, $c_{2\beta}=\cos 2\beta$, $s_{2\beta}\equiv\sin2\beta$  and $v=\sqrt{v_u^2+v_d^2}=174$ GeV and we have omitted the entries below the diagonal for simplicity. 
This matrix is related to the interaction eigenstate mass matrix by a $t_\beta$ dependent rotation, and we have re-written the parameter $A_\lambda$ in  terms of the usual MSSM parameter $m_A$ as follows:
\begin{equation}
m_A^2=\frac{\mu}{s_\beta c_\beta} \left(A_\lambda +
\frac{\kappa\mu}{\lambda} \right). \label{eq:mAAlambda}
\end{equation}  
In the absence of significant mixing between the different states, the mass of $H$ will be approximately given, as in the MSSM, by the mass parameter $m_A$.

The problematic element of this matrix is the off-diagonal $h-S$ term:
since $m_A$ must be kept fairly large in order to lift the heavy 
CP-odd/even masses in accordance with LHC constraints, this mixing term tends
to be large, leading to a tachyonic eigenvalue upon diagonalization.  Additionally, this
mixing can induce sizable deviations of the SM-like Higgs couplings,
rendering it non-SM-like. This off-diagonal term can, however, be tuned
away by choosing parameters such that $m_A^2 \sim 4 \mu^2
/ s_{2\beta}^2$.  Additional $h-S$ mixing is induced through the
off-diagonal $h-H$ and $H-S$ terms, but this is hierarchically smaller than
mixing induced directly by the off-diagonal $h-S$ term and generically
evades LHC bounds.  Both $h-S$ and $H-S$ mixing also induce scattering in
direct detection experiments, which are generically sizable for points
where the neutralino couples strongly enough to produce the GCE.

Likewise, the CP-odd mass matrix, in the $(A,S)$ basis is
\begin{eqnarray}
\mathcal{M}^2_{P} & = & \left(
\begin{array}{cc}
m_A^2 &\lambda v \left(\frac{m_A^2 }{2\mu }s_{2\beta}-\frac{3
  \kappa \mu }{\lambda }\right) \\
& \lambda ^2 v^2 s_{2\beta} \left(\frac{m_A^2}{4\mu ^2} s_{2\beta}+\frac{3 \kappa }{2\lambda }\right)-\frac{3 \kappa A_{\kappa }
  \mu }{\lambda }\\
\end{array} 
\right),\nonumber\\
\end{eqnarray}
where $A$ denotes the MSSM pseudoscalar in the absence of the singlet. The lighter and heavier mass eigenstates will be denoted by $m_a$ and $m_{a_2}$ respectively. Note that in the presence of significant mixing between the two states, the mass of the heavier state, $m_{a_2}$ can be quite discrepant from the MSSM pseudoscalar mass parameter, $m_A$. Generically such significant mixing will exist between the singlet and MSSM-like component of the lightest
pseudoscalar eigenstate; there is insufficient freedom to remove this
mixing in the $Z_3$ NMSSM, though it may exist in the full NMSSM.
Further, by choosing $A_\kappa$, $m_a$  can be tuned to
desirable values as needed for annihilating $2 \rightarrow 2$ through
the light CP-odd pseudoscalar.

Our results can be summarized as follows:
\begin{itemize}
\item For mixed Singlino/Higgsino dark matter, annihilation is
  mediated via the pseudoscalar on resonance in the GC today, while the
  relic abundance is set by a combination of annihilation through the
  pseudoscalar and the $Z$ boson. We will show that off-resonance
  annihilation is not possible in this case on account of the $Z$ pole: our analysis in
  Sec.~\ref{Sec:Simplified} shows that a large product of couplings is
  necessary, implying a large value of $\lambda$ and correspondingly large Higgsino fraction.  At large $t_\beta$, this enhances the $Z$ contribution to
  the relic density and may violate $Z$-pole constraints; at small $t_\beta$ this produces a sizable direct detection
  cross-section which cannot be tuned away.  In either case, the constraints on the Higgsino fraction force annihilation near the pseudo scalar resonance.
\item For mixed Bino/Higgsino dark mater, contrary to the S/H case,
  annihilation must occur off-resonance, unless $\mu$ is very
  large. We will show that the needed hierarchy of couplings for resonant annihilation discussed in Fig.~\ref{FreezeoutAndGC} cannot be achieved for the B/H case.
\end{itemize}

\subsection{Singlino/Higgsino Dark Matter $(\kappa/\lambda \ll 1)$}\label{sec:SH}

We begin by expanding the components of the neutralino in the limit $\kappa/\lambda \ll 1$, so that we can read off the coupling of the DM to the (mostly singlet) CP-odd scalar which mediates the annihilation. The full expressions can be found in Appendix~\ref{Ap:Neut}. We find  
\begin{eqnarray}
\frac{N_{13}}{N_{15}} &\sim& - \frac{v \lambda}{\mu} c_\beta \left(1 - t_{\beta} \frac{m_{\chi}}{\mu}\right),~~~\frac{N_{14}}{N_{15}} \sim - \frac{v \lambda}{\mu} s_{\beta}\left(1 - \frac{m_{\chi}}{\mu t_{\beta} }\right), \label{HiggsinoComponents} \\
 N_{15} &\sim& \left[ 1 + \frac{v^2 \lambda^2}{\mu^2}\left(1 + s_{2\beta} \frac{m_{\chi}}{\mu}\right)\right]^{-1/2}, 
\end{eqnarray}
where $m_{\chi}/\mu$ is also taken to be small, and $N_{13}$, $N_{14}$ and $N_{15}$ refer to the Higgsino down, Higgsino up and Singlino components of the lightest neutralino respectively. 

In the S/H scenario, the SM-like Higgs can mix strongly with the light singlet-like Higgs. We will always assume this mixing is suppressed since it leads to non-SM like behavior for the 125 GeV Higgs. As detailed in Appendix~\ref{sec:CPE}, this forces $m_A \sim 2 |\mu|/s_{2\beta}$ which removes the possible MSSM type $t_\beta$ enhancement one could expect for the coupling of the pseudoscalar to the down type quarks. 

The annihilation of a pair of neutralinos via a pseudoscalar proceeds predominantly to $b\bar b$, so that the relevant quantity of interest is the active part of the mostly singlet pseudoscalar. Assuming that $m_A >> m_a$, this component is given by
\begin{equation}
\frac{P_{a,A}}{P_{a,S}} \sim - v \frac{\lambda \, s_{2\beta}}{2 \mu}.
\end{equation}
where $P_{i,j}$  indicates the composition of pseudoscalar mass eigenstate $i$ ($i=a,a_2$), in terms of the interaction eigenstates $j$ ($j=A, S$). 
The generally larger singlet component of the lightest pseudoscalar is upon normalization:
\begin{equation}
P_{a,S}\sim\left(1+\frac{\lambda^{2}v^{2} s_{2\beta}^2}{4 \mu^{2}} \right)^{-1/2}. \label{eq:PasNormSH}
\end{equation}
We thus find that the couplings of the lightest pseudoscalar to the DM and the $b$ quarks can be written 
\begin{eqnarray}
g_{a \chi \chi} & \sim & i \sqrt{2} \left[\kappa N_{15}^2 -\lambda  N_{13} N_{14}+\frac{\lambda ^2 v}{2 \mu } s_{2\beta } \left(N_{13} c_\beta+N_{14} s_\beta\right) N_{15}\right] P_{a,S},   \label{gaxx} \\ 
g_{a b b} & \sim & -i \frac{m_b \lambda}{\sqrt{2}\mu} s_\beta^2P_{a,S}, \label{gabb}
\end{eqnarray}
where one can see that there is no $t_\beta$ enhancement in the couplings. This implies that, as one moves away from resonance, $\lambda/\mu$ will have to grow substantially to maintain the required annihilation rate for the GCE. 

We also see from Eq.~\ref{HiggsinoComponents} that the Higgsino
component may be substantial (unless $\mu$ is very large).  This
generates a coupling to the $Z$-boson of
\begin{equation}
g_{Z\chi\chi}=\frac{m_Z}{\sqrt{2} v}\left(N_{13}^2-N_{14}^2\right), 
\end{equation}
which vanishes in the limit of $t_\beta \rightarrow 1$.  Because twice
the mass of the DM in the $2 \rightarrow 2$ annihilation is close to
$m_Z$, annihilation through the $Z$ pole is important for setting
the relic abundance away from $t_\beta = 1$. On the other hand, since
annihilation of a Majorana particle through a vector particle is
$p$-wave suppressed, this annihilation mode is unimportant in the
Universe today. We verified that there is no destructive interference
between the $Z$ and a possibly resonant (though $p$-wave suppressed)
annihilation via the singlet-like scalar. Therefore, to obtain a GCE,
we need the $Z$ mediated thermal relic density to not be too large.

\begin{figure}[tbh]
\includegraphics[width=0.48\textwidth]{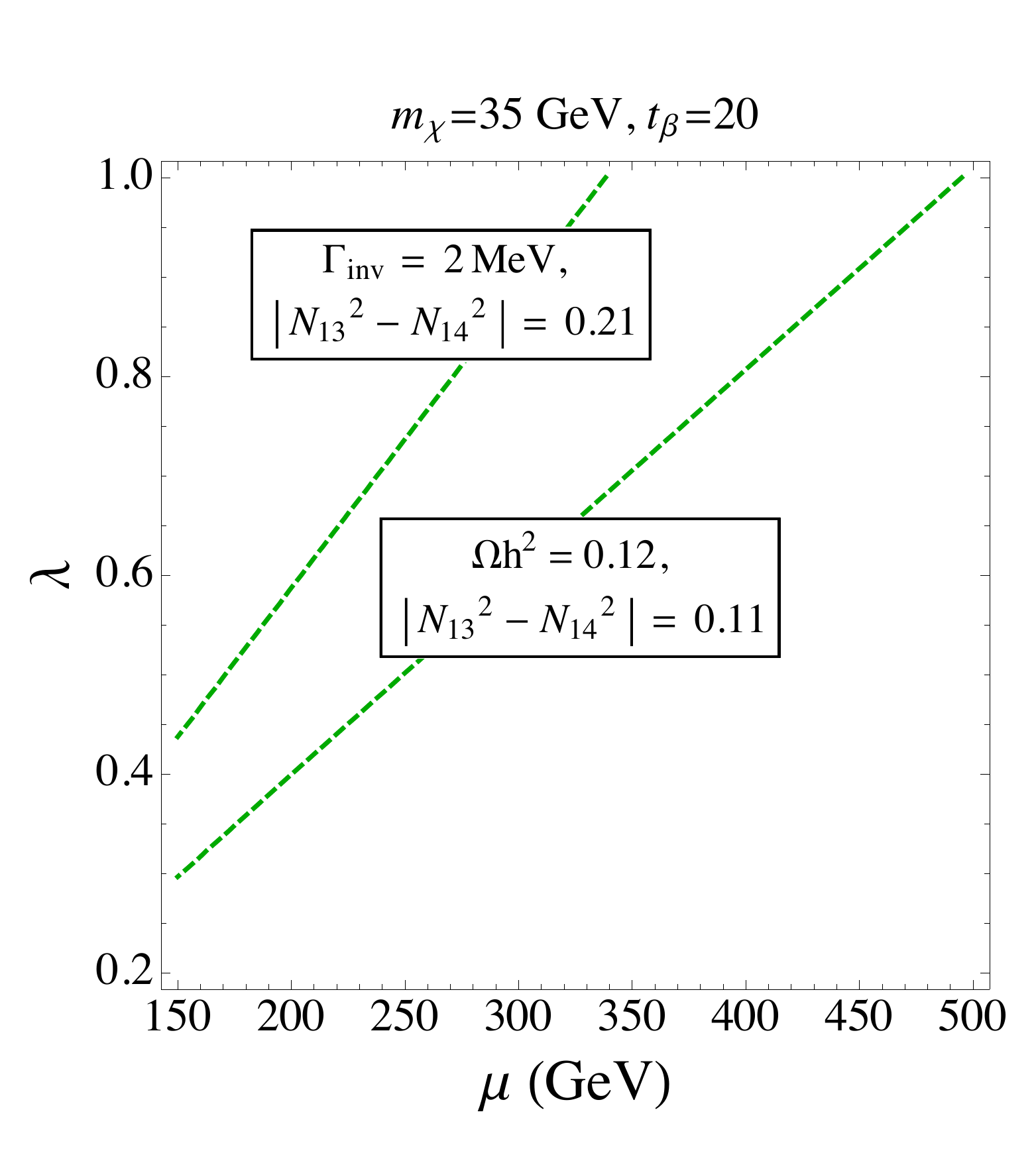} 
\includegraphics[width=0.48\textwidth]{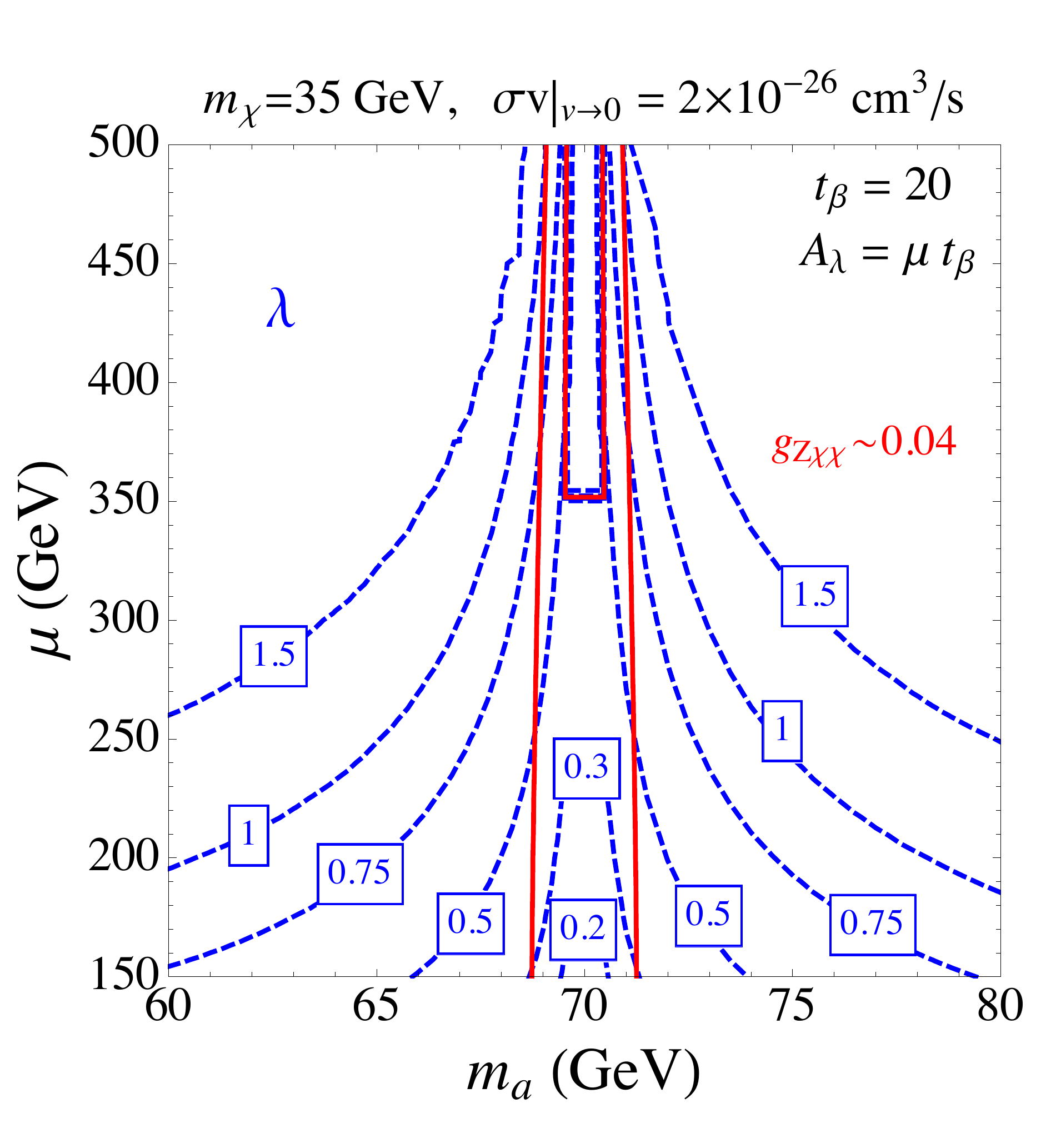} 
\caption{{\em Left:} $\lambda$ as a function of $\mu$ needed to set the relic density via annihilation through the $Z$.  The relic abundance thus fixes an upper bound on the Higgsino fraction. {\em Right:} Blue~(dashed) curves denote the value of $\lambda$ needed to obtain the GCE in the $\mu$ versus pseudoscalar mass, $m_a$, plane.  The red~(solid) curve traces out where the Higgsino fraction is such that one  obtains the correct relic abundance via annihilation through the $Z$ alone.  As one can see, annihilation must occur very close to resonance to achieve the GCE under these conditions.}
\label{2->2SinglinoHiggsino}
\end{figure}

We used {\tt micrOMEGAs} to obtain the value of $g_{Z\chi\chi}$ leading to the observed thermal relic density for $m_\chi =35$ GeV via annihilation through the $Z$ pole: $g_{Z\chi\chi} \sim$ 0.04.
The
contour corresponding to this coupling is shown in the $\lambda$ -
$\mu$ plane for $t_\beta=20$ in the left panel of
Fig.~\ref{2->2SinglinoHiggsino}, setting an upper bound on $\lambda$
for a given value of $\mu$.

The invisible width of the $Z$ gives another constraint on $g_{Z\chi\chi}$.  The partial width of the $Z$ to a pair of neutralinos is
given by
\begin{equation}
\Gamma  = \frac{G_F m_Z^3}{ 12\sqrt{2} \pi} \left(N_{13}^2-N_{14}^2\right)^2\left(1-\frac{4 m_\chi^2}{m_Z^2}\right)^{3/2},
\end{equation}
and is constrained to be $\lesssim~2~\mev$~\cite{ALEPH:2005ab}, yielding $|N_{13}^2-N_{14}^2|\left(1-4
m_\chi^2/m_Z^2\right)^{3/4} \lesssim 0.11$. This upper bound is also
shown in Fig.~\ref{2->2SinglinoHiggsino} as a function of $\mu$. Since
the dependence of the neutralino composition upon $\lambda$ and $\mu$ does not
change significantly when one extends the $Z_3$ NMSSM to the general
NMSSM~(See Appendix.~\ref{Ap:Neut}), this requirement extends robustly
to the general NMSSM.  While the upper bound from $\Omega h^2 \sim
0.12$ is more constraining for $m_\chi = 35~\gev$, the bound from the
invisible width of the $Z$-boson becomes more constraining for lighter
dark matter masses, due to the phase space opening.

We further extract the needed $\lambda$ to obtain the GCE for a given
pseudoscalar mass $m_a$; this is shown in the right panel of
Fig.~\ref{2->2SinglinoHiggsino}. Given that the coupling of the $Z$
boson and therefore its contribution to the relic density are fixed by
the Higgsino component of the DM (and hence by $\lambda$ and $\mu$),
we can see from the right panel of Fig.~\ref{2->2SinglinoHiggsino}
that the combination of the GCE plus relic abundance implies that
$m_{a}$ is very close to $2m_{\chi}$: Outside the region denoted by
the red~(solid) curve, the thermal cross-section from the annihilation
via $Z$ alone would force a too small relic density, so that there
cannot be any additional contribution from the annihilation via the
pseudoscalar.  The internal $g_{Z\chi\chi} \sim 0.04$ line present for
$\mu \gtrsim 350$ in the right panel is the same solution as the
external pair of lines, and results from the modified values of
$\lambda$ need to achieve the GCE near-resonance.

Even on-resonance, to be phenomenologically viable, the spin-independent direct detection cross-section must evade the stringent LUX bounds for $m_\chi\sim 35$ GeV, $\sigma_{SI} \lesssim 10^{-9}$ pb. Extracting the SM-like Higgs only  contribution  from the general expression presented in Appendix \ref{sec:DD}, the scattering cross-section is:
\begin{eqnarray}
\sigma_{SI}^{h} & \simeq &\lambda^4 \frac{m_r^2}{\pi m_h^4} \left[\frac{m_p(\mu ~s_{2\beta}-m_\chi)}{\mu^2-m_\chi^2}\right]^2 N_{15}^2 \left(\sum_{q=u,d,s} f_{Tq} + \frac{6}{27} f_{TG}\right)^2 \\
& \simeq & 1.2 \times 10^{-45} \mbox{ cm}^2 \times N_{15}^2 \left(\frac{\mu ~s_{2\beta}-m_\chi}{\mu-m_\chi}\right)^2 \left(\frac{\lambda}{0.2}\right)^4\left(\frac{200 \mbox{ GeV}}{\mu+m_\chi}\right)^2 \left(\frac{125 \mbox{ GeV}}{m_h}\right)^4. \nonumber \\
\label{SMHiggsDD}
\end{eqnarray}
While this scattering may be small when $t_\beta$ is large, when $t_\beta = 1$, this scattering cross-section is generally above current bounds.

Depending on parameters, however, destructive interference can render the spin-independent scattering cross-section small, and even vanishing.  For example, this can occur for neutralino-DM scattering in the MSSM.  Even if the only exchanged particle is the Higgs, depending on the admixture of Bino and Higgsino in the DM, the scattering cross-section can identically vanish, {\it i.e.}~there may be a direct detection blind spot~\cite{Cheung:2012qy}.  If multiple MSSM scalars exist in the spectrum, there may also be destructive interference due to multiple scalar exchange channels~\cite{Huang:2014xua}.  A systematic study of blind spots in the NMSSM does not exist, although blind spots in a broader class of simplified DM models were considered in~\cite{Cheung:2013dua}.  In the present scenario, blind spots may result from destructive interference among the exchanged scalar states.  Combining the results in Appendix \ref{sec:DD}, for moderate/large $t_\beta$ we have
\begin{eqnarray}
\sigma_{SI} &\simeq& \frac{ m_p^2 m_r^2}{v^2\pi } 
\left\{\frac{\left(F_d+F_u\right)}{ m_h^2 t_{\beta }} 
  \left[\lambda  N_{13} t_{\beta } \left(N_{14} S_{h,s}-N_{15}\right)
  -N_{15} \left(\lambda  N_{14}+\kappa  N_{15} S_{h,s} t_{\beta }\right)\right] \right. \nonumber \\
   && +\frac{\left(F_d t_{\beta }^2-F_u\right) }{ m_H^2 t_{\beta}^2} 
   \left[\lambda  N_{13} \left(N_{14} S_{H,s} t_{\beta}+N_{15}\right)
   -N_{15} t_{\beta } \left(\lambda  N_{14}+\kappa  N_{15} S_{H,s}\right)\right] \nonumber \\
  &&   -\frac{\left(F_d t_{\beta } S_{h_{S},d}+F_u S_{h_S,u}\right)}{ m_{h_S}^2} 
  \left[N_{15} \left(\lambda  N_{14} S_{h_{S},d}-\kappa  N_{15} S_{h_{S},s}\right) \right. \nonumber \\
   && \qquad \qquad \qquad  \left. \left.+\lambda 
   N_{13} \left(N_{15} S_{h_S,u}+N_{14} S_{h_{S},s}\right)\right]\right\}^2, \label{eq:SHDDtb}
   \end{eqnarray}
where $F_u = \sum_{q=u} f_{Tq} + \frac{4}{27} f_{TG} \sim 0.15,~F_d = \sum_{q=d,s} f_{Tq} + \frac{2}{27} f_{TG} \sim 0.13$.  This allows for a 3-way cancellation between the contributions from $h_S,h,H$, as we will show below.

At large/moderate $t_\beta$, the small up and down components of the singlet-like Higgs are related to the singlet components of the standard and non-standard heavy Higgs by:
\begin{eqnarray}
S_{h_{S},d} &\sim & S_{H,s}+\frac{S_{h,s}}{t_{\beta }}, \label{eq:sd}\\
S_{h_{S},u} &\sim& S_{h,s}-\frac{S_{H,s}}{t_{\beta }},  \label{eq:su}\\
\text{where        }\qquad S_{H,s} &\sim& \lambda v/\mu t_\beta. \label{eq:snsm}
\end{eqnarray}
As mentioned previously, the singlet component of the SM-like Higgs can be minimized by tuning $m_A \sim \mu t_\beta$, though this relationship receives relevant radiative corrections which can introduce a non-zero (though small) mixing angle: 
\begin{eqnarray}
S_{h,s} &\approx& \frac{-2\lambda  v \mu \epsilon  }{(m_{h}^2-m_{h_S}^2)}, 
\end{eqnarray}
where $\epsilon$ parametrizes the departure of this mixing angle from the tree-level cancellation induced by setting 
\begin{equation}
m_A^2 =\frac{ 4\mu ^2}{s^2_{2\beta}} \left(1-\frac{\kappa }{2\lambda } s_{2\beta}-\epsilon\right)|_{\epsilon\rightarrow 0}. \label{eq:epsilon}
\end{equation}
The singlet-SM-like Higgs mixing is also relevant for the singlet-like Higgs mass, which can be approximated by:
\begin{equation}
m_{h_S}^2 \sim \frac{\mathcal{M}^2_{hS}(2,2)+\delta _{\text{loop}}-m_h^2
   \left(1-S_{h,s}^2\right)
   S_{h,s}^2}{1+\left(1-S_{h,s}^2\right) S_{h,s}^2}
\end{equation}
where 
\begin{eqnarray}
\mathcal{M}^2_{hS}(2,2)&=&\frac{\kappa  \mu 
  }{\lambda } \left(A_{\kappa }+\frac{4 \kappa  \mu }{\lambda }\right)+\lambda ^2 v^2 \left(1-c_{2 \beta }^2\right)-\frac{\kappa ^2  v^2 }{2} s_{2\beta}^2 \,c_{2 \beta
   }^2-\frac{1}{2} \kappa  \lambda  v^2 \left(2 c_{2 \beta }^2+1\right) s_{2 \beta } \nonumber\\
   \label{eq.mhs}
\end{eqnarray}
and the dominant contribution to $\delta_{\text{loop}}$ is~\cite{Draper:2010ew, Carena:2011jy}
\begin{equation}
\delta_{\text{loop}} \sim \frac{\lambda^2\mu^2}{2\pi^2}\log \frac{m_H^2}{\mu^2} \sim \frac{\lambda^2\mu^2}{2\pi^2}\log t_\beta^2 \, .
\end{equation}

Using Eqs.~\ref{HiggsinoComponents} and \ref{eq:sd}-\ref{eq:snsm} in Eq.~\ref{eq:SHDDtb}, the direct detection cross-section is then proportional to (again in the large $t_\beta$ limit)
\begin{eqnarray}
\sigma_{SI} & \propto & \left\{\left(\frac{2}{t_{\beta}}-\frac{m_{\chi}}{\mu}\right)\frac{2\, t_{\beta}}{m_h^2}
 +\frac{t_{\beta }}{m_H^2} \right.  \nonumber \\
 & + &    \left.\frac{1}{m_{h_S}^2}\left(2\, S_{h,s}+\frac{\lambda\,  v}{\mu}\right) 
 \left[\frac{\lambda  \,v}{\mu^{2} }\, m_{\chi }  +S_{h,s}\left(\frac{2}{t_{\beta}}-\frac{m_{\chi}}{\mu}\right)
 +\frac{\kappa \, \mu }{\lambda^{2} \,v}\right] 
 \right\}^2. \nonumber \\ \label{eq:SHcanceltb}
\end{eqnarray}
We can see from the above that positive values of $\mu$ lead to suppression of the spin-independent direct detection cross-section. First, we note that $\mu >0$  has the effect of reducing the Higgsino component and therefore the dominant contribution due to the SM-like Higgs. Second, since $m_H\sim m_A \sim |\mu| \,t_\beta$, the direct detection cross-section is further reduced when $\mu$ is positive. Therefore, generally, direct detection bounds do not constrain very strongly the region of interest in the $\lambda$-$\mu$ plane,

\begin{figure}[!h]
\begin{tabular}{cc}
\includegraphics[width=0.48\textwidth]{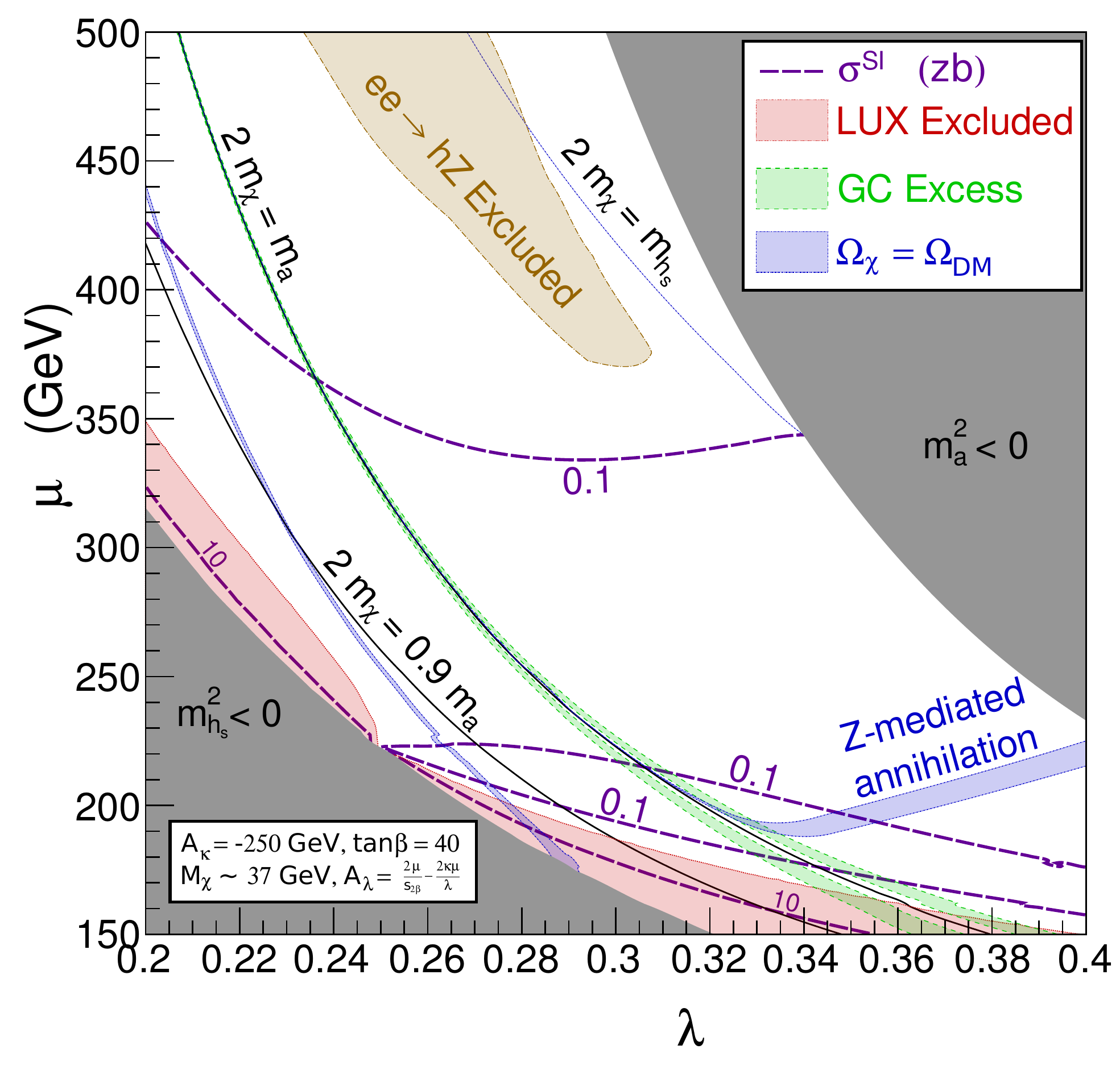} &
\includegraphics[width=0.48\textwidth]{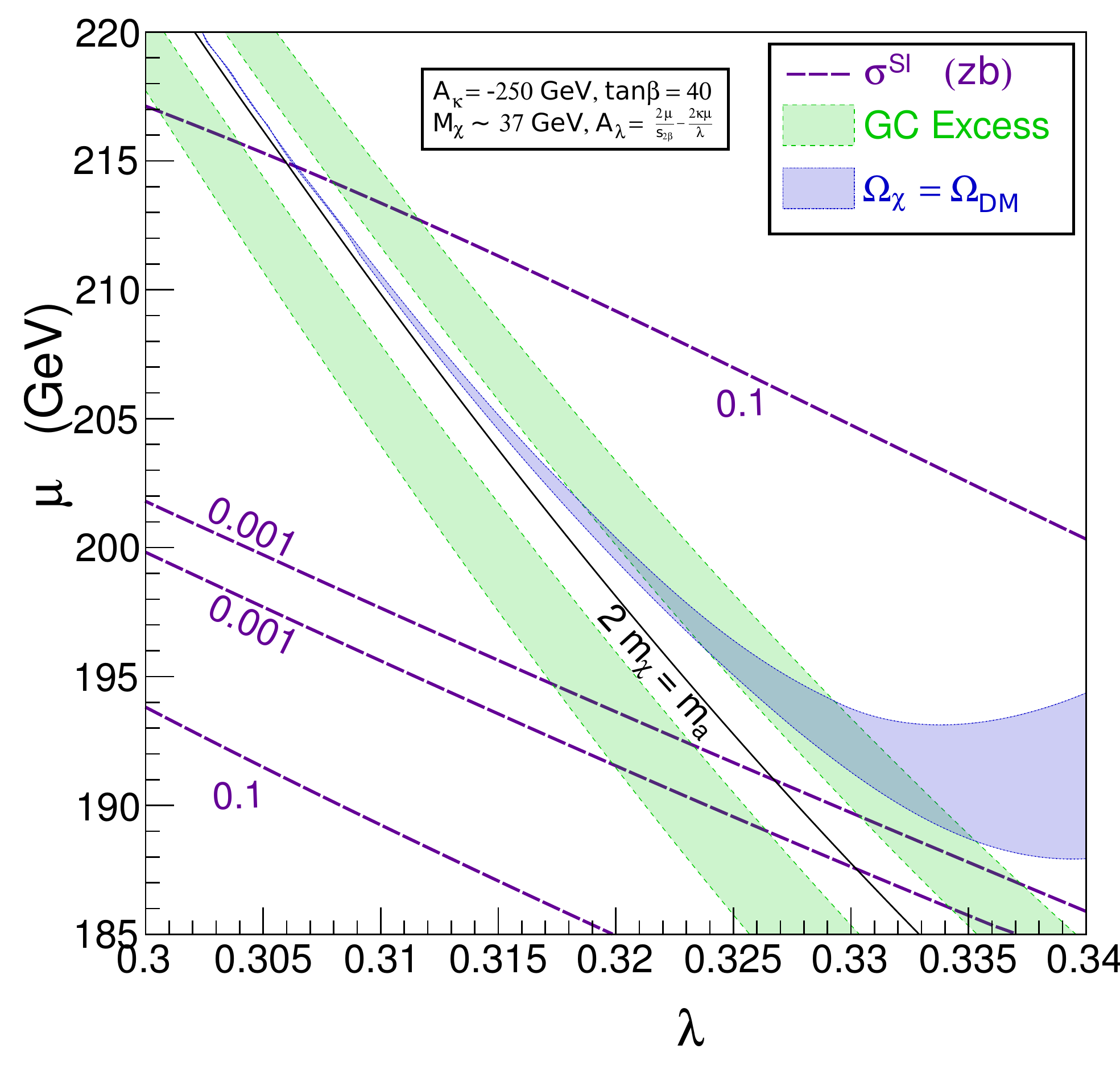} \\
\includegraphics[width=0.48\textwidth]{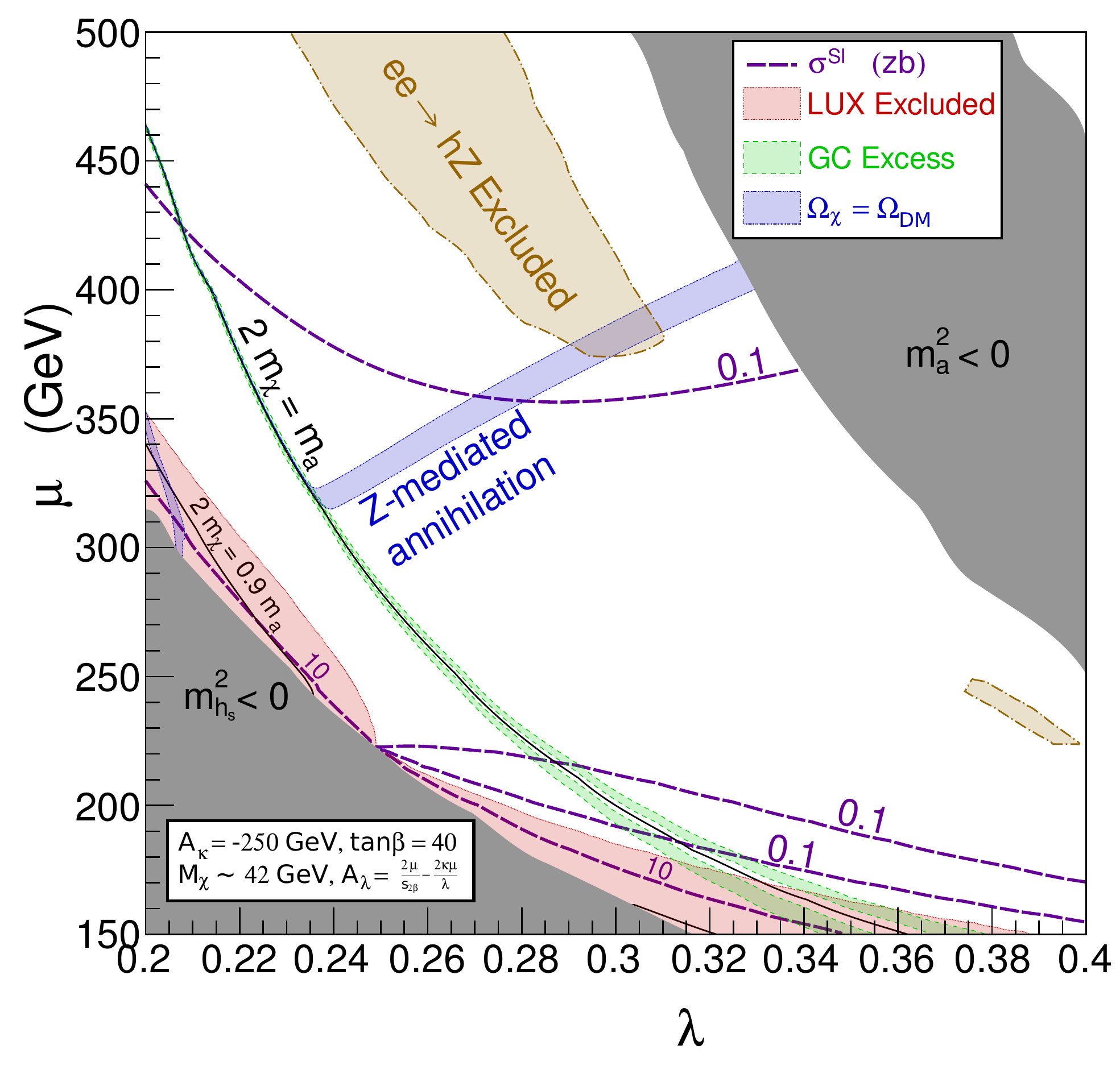} &
\includegraphics[width=0.48\textwidth]{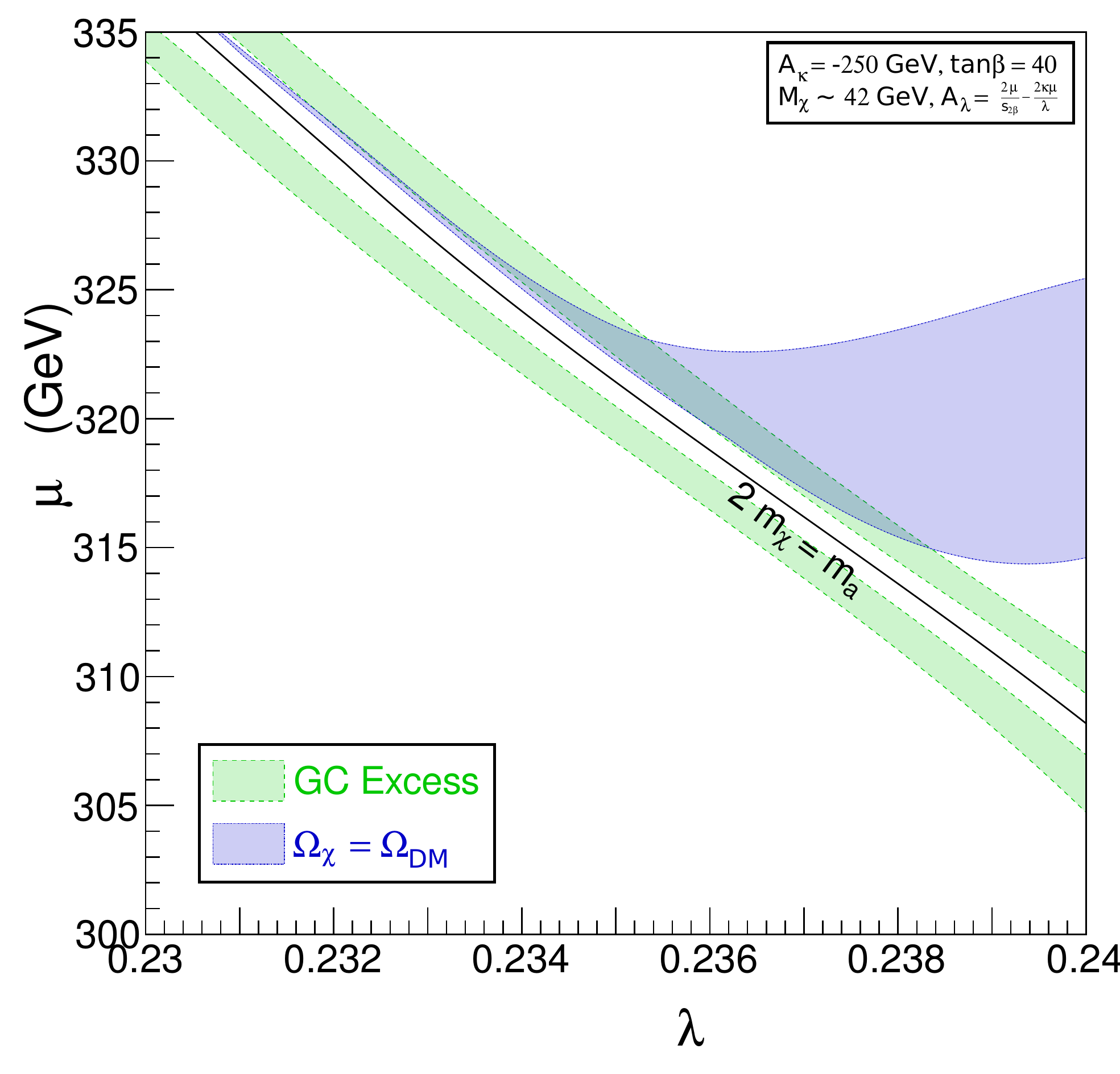} 
\end{tabular}
\caption{Results of a numerical scan with {\tt micrOMEGAs}, fixing $m_h \sim$ 125 GeV,
$t_\beta = 40$, $A_\kappa=-250$ GeV, and $A_\lambda$ to remove mixing between the SM-like and singlet Higgses.  The green band shows the region of parameter space fitting the GCE: $0.5 <  10^{26}$ cm$^3$/s$ \times \sigma v|_{v\rightarrow 0} <$  4, the blue region shows  the observed relic abundance, and the red the excluded LUX region.  We have fixed $\kappa$ to accommodate particular values of the DM mass. {\em Upper panels}: $m_\chi \sim 37\pm0.5$ GeV, {\em Lower panels}: $m_\chi \sim 42\pm0.5$ GeV.   Consistent with the analytic results shown in Fig.~\ref{2->2SinglinoHiggsino}, the green strips are centered around $m_a = 2 m_\chi$, and as $\lambda/\mu$ (controlling the Higgsino fraction) decreases, the green strips converge closer to resonant annihilation. The blue relic density strip breaks away from the green GCE line when annihilation through the $Z$ becomes important.}
\label{2->2SHMO}
\end{figure}

To verify our analytics, we performed a numerical scan in the NMSSM parameter space using {\tt NMSSMTools 4.2.1}, which in turn runs {\tt micrOMEGAs\_3.0}. The results are summarized in Fig.~\ref{2->2SHMO}.  A priori, each point in our parameter space is defined by the six parameters: $(\lambda, \kappa, A_\lambda, A_\kappa, \mu, t_\beta)$.  In Fig.~\ref{2->2SHMO}, $\lambda$ and $\mu$ are plotted as axes.  At each point we have fixed $t_\beta=40$, $A_\kappa=-250$ GeV, and adjusted $A_\lambda$ to set $\epsilon$ to zero at tree-level (Eq.~\ref{eq:epsilon})~\footnote{We verified that this condition renders the 125 GeV Higgs very SM-like.}. As a result, singlet-Higgs mixing is only generated from loop effects, and is thus small.   For the top and bottom panels, we have fixed $\kappa$ to accommodate a DM mass of $\sim$ 37 GeV and $\sim$ 42 GeV, respectively, using the tree level relation in Eq.~\ref{eq:kappa}. The right panels show the zoomed in region of interest for the corresponding DM mass.

At large $t_\beta$, the NMSSM coupling $\lambda$ does not help to
boost the Higgs mass.  Consequently, we require a heavy stop sector to
lift the Higgs, as in the MSSM.  Thus, at each point in
Fig.~\ref{2->2SHMO} we have fixed the stop masses such that $m_h =
125$ GeV, with $A_t = 0$.  Both the singlet-like scalar and
singlet-like pseudoscalar masses vary in this plane, and the gray
shaded regions denote where one or the other becomes tachyonic, in
which case there is no successful electroweak symmetry breaking. The
solid black lines in Fig.~\ref{2->2SHMO} show contours where
$2m_{\chi} = \{m_a, 0.9\, m_a\}$.

The green shaded region denotes the region roughly consistent with the
GCE, with $0.5 \times 10^{-26} \textrm{ cm}^3/\textrm{s}< \sigma
v|_{v\rightarrow 0} <4\times 10^{-26} \textrm{ cm}^3/\textrm{s}$.  As
expected from our analytical results, this region is composed of two
distinct strips which closely straddle the contour $2
m_\chi/m_a=1$.  These strips are relatively wide and further from
resonance for lower values of $\mu$, and become narrower and closer to
resonance for larger values of $\mu$: both the decrease in the
Higgsino fraction and the decrease in $\lambda$ produce a reduction in the
$a\chi\chi$ coupling for larger values of $\mu$, thus requiring more
resonant behavior to produce sufficient annihilation.

The blue shaded regions in the plot denote where the relic density is
consistent with the experimentally observed one within $\pm 3\sigma$:
0.1118 $ < \Omega h^2 <$ 0.128.  Like the GCE, there are two blue
bands, corresponding to either side of the resonant point.  The blue
ribbon that breaks away from the resonant region and follows
constant $\lambda/\mu$ denotes where the relic density is controlled
by annihilation via the $Z$.  For $m_\chi\sim 37$ GeV, this is in good agreement with what is
shown for $m_\chi=35$ GeV in the left panel of Fig.~\ref{2->2SinglinoHiggsino}. Further, for
$m_\chi\sim 37~\gev$ there is a small sliver of parameter space where
the relic density is achieved through resonant annihilation via the
CP-even scalar.  This region is not viable for the GCE because this
process is $p$-wave suppressed in the present day.

As we saw in our discussion of the simplified model, the regions consistent
with a thermal relic and GCE are disjoint if the only annihilation
process is via $s$-channel pseudoscalar exchange and a delicate
balance between the couplings and masses is required to make both GCE
and relic density consistent at the same time. The green and blue
bands are very different in the region $2m_\chi < m_a$.  The relic
density band is located at $2 m_\chi \sim 0.9\, m_a$, as thermal
averaging allows resonant annihilation in the early universe, while
the GCE region must be close to resonance for any enhancement in the
annihilation rate.

Meanwhile, the green and blue bands for $2m_\chi >m_a$ appear
coincident, but do not overlap even when both appear to merge into the
$2m_\chi >m_a$ contour.  The reason is again the presence of thermal
broadening in the early universe, which in this case results in more
neutralinos annihilating off-resonance and thus pushes the relic line
closer to resonance.  However, in the presence of a second
annihilation channel---in particular, via the $Z$ boson---this changes
because another relic annihilation mode is in play.  Consequently, the blue and
green bands do cross once the $Z$ contribution starts to matter.  In
the left and right panels of Fig.~\ref{2->2SHMO}, the thermal relic
and GCE bands overlap at $(\lambda, \mu) \sim (0.33, 200\textrm{ GeV})$ and
$(\lambda, \mu) \sim (0.24, 320\textrm{ GeV})$, respectively.  The difference
in the values of $\lambda$ and $\mu$ comes from the slight difference in the DM masses
between the two panels.  In the right panel, the DM mass
is slightly larger, and is thus closer to the $Z$ boson resonance,
requiring a larger value of $\mu$ to suppress the cross-section to
appropriate levels.

We checked that all these points are in agreement with the recent LHC limits on chargino
neutralino direct production. We find $BR(\chi^{0}_{2}\rightarrow\chi^{0}_{1}Z)\sim 0.4$ with {\tt micrOMEGAs} for the region of interest. On the other hand, by taking the ATLAS trilepton search~\cite{Aad:2014nua} and using the provided simplified model
information on $\chi^{0}_{2}\chi^{\pm}_{1}\rightarrow W^{\pm}Z\chi^{0}_{1}\chi^{0}_{1}$, we found that for $\mu\gtrsim 150\,{\rm GeV}$ the upper limit on $BR(\chi^{0}_{2}\rightarrow\chi^{0}_{1}Z)$ is always weaker than $0.4$ (which is reached only at $\mu\sim 200\,{\rm GeV}$) for the values of DM mass considered here.

The red shaded regions in Fig.~\ref{2->2SHMO} are excluded by LUX. We
see clearly that direct detection does not provide a very stringent
constraint on the parameter space.  For smaller values of $\mu$ there
is a blind spot in the parameter space, at which the SI direct
detection cross-section vanishes identically.  At larger values of
$\mu$, meanwhile, the SI cross-section falls off.  We have verified
that the region where the direct detection cross-section is minimized
is in very good agreement with that predicted by
Eq.~\ref{eq:SHcanceltb}.  Spin-dependent direct detection results from
XENON100~\cite{Aprile:2013doa} do not currently place constraints on
the parameter space shown, but LUX spin-dependent results may
constrain larger values of $\lambda$.

In principle, one could vary $A_\kappa$ for any set of parameters to independently set the pseudoscalar mass via Eq.~\ref{eq:akappa}. However, we want to show the variation in the cosmological quantities of interest with the pseudoscalar mass. Taking a larger (smaller) value of $A_\kappa$ would shift the relic density and GCE contours to the right (left), leading to somewhat larger (smaller) values of $\lambda$ and $\mu$.

\begin{figure}[t]
\includegraphics[width=0.48\textwidth]{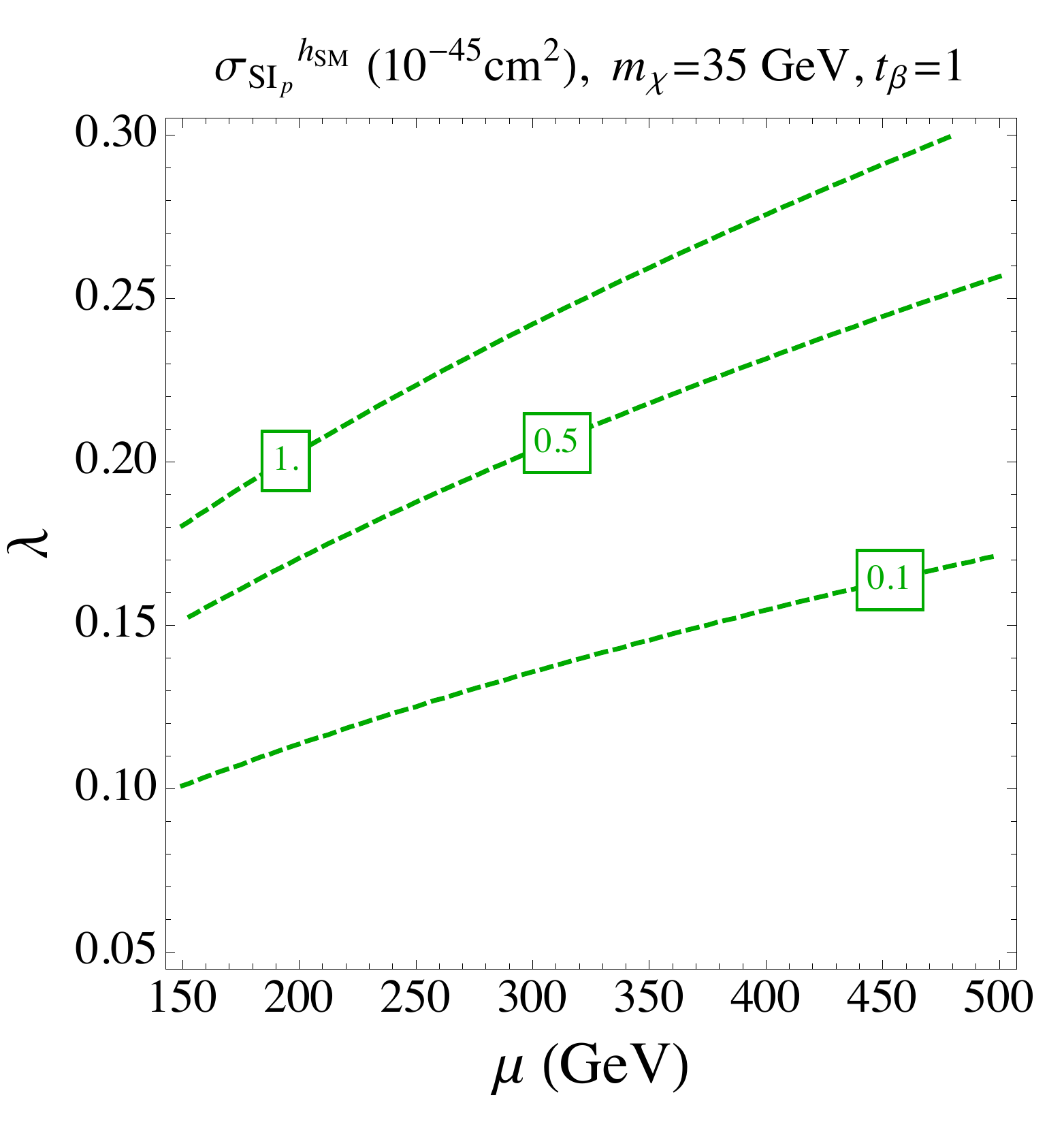} 
\includegraphics[width=0.48\textwidth]{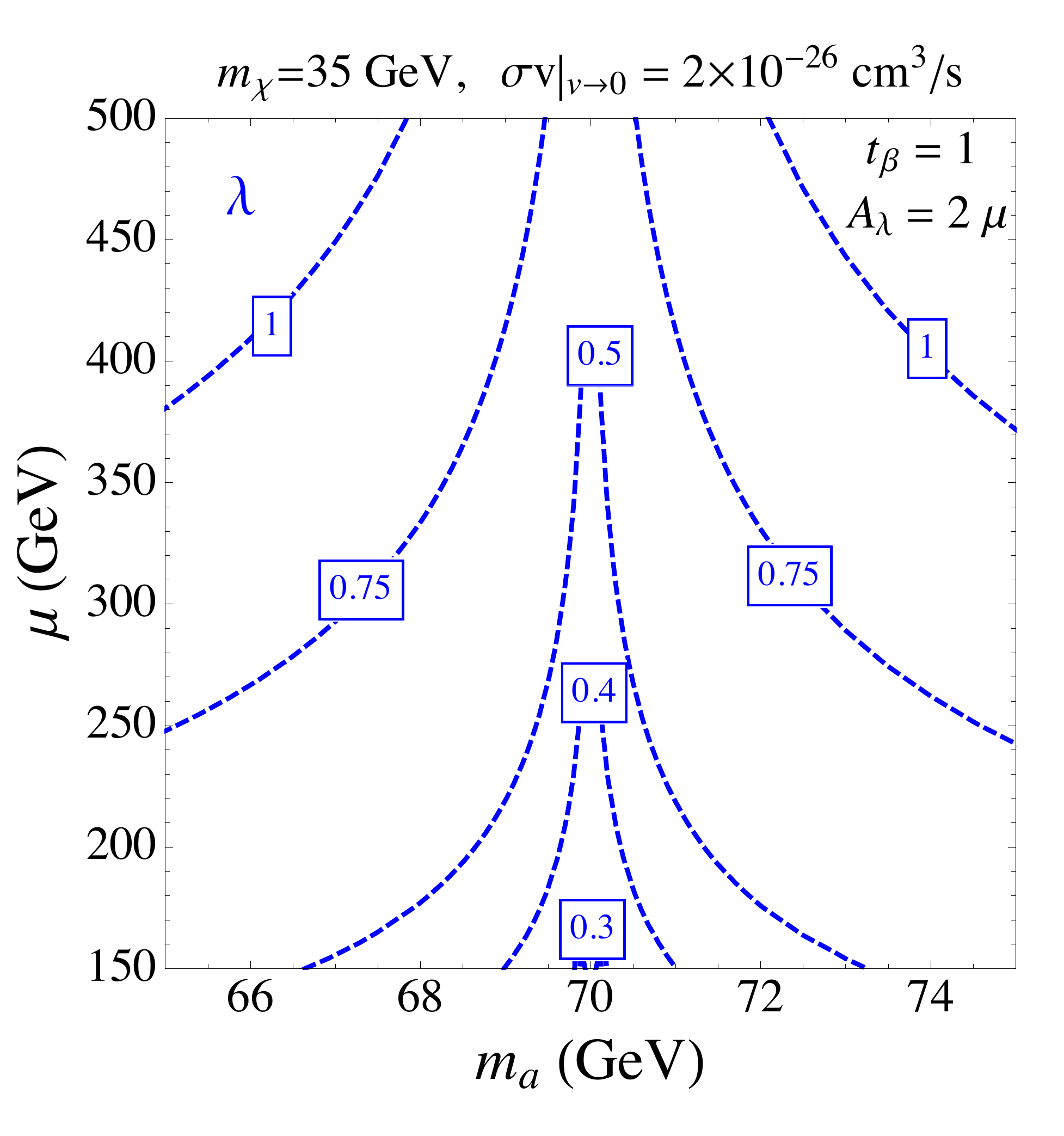} 
\caption{{\em Left: } Contours of direct detection cross-section (in
  units of $10^{-45} \mbox{ cm}^2$) in the $\lambda-\mu$ plane, taking
  the analytic expressions in Eq.~\ref{SMHiggsDD} and $\tan \beta =
  1$.  {\em Right: } Contours of $\lambda$ needed to obtain the
  galactic center excess.  Comparing with the left-hand plot, we can
  see that the annihilation must occur very close to resonance.}
\label{2->2SinglinoHiggsinotb1}
\end{figure}

We now consider the case of $t_\beta=1$. Here, there is no
contribution from the $Z$ to the relic density and in principle one
could obtain both consistent GCE and relic density just from the
exchange of the pseudoscalar. However, in the absence of blind spots,
the direct detection cross-section is large through the SM-like Higgs,
as can be seen in Eq.~\ref{SMHiggsDD} and in the left panel of
Fig.~\ref{2->2SinglinoHiggsinotb1}. The right panel, similar to the
right panel of Fig.~\ref{2->2SinglinoHiggsino} shows the required
values of $\lambda$ in the $m_a$-$\mu$ plane to obtain the
GCE. Comparing the left and right panels of
Fig.~\ref{2->2SinglinoHiggsinotb1}, we can see that without a
significant reduction of the direct detection cross-section from the
SM-like Higgs, off-resonance annihilation would be ruled out by direct
detection constraints.

There is no contribution from the heavy MSSM like non-standard Higgs to the direct detection cross-section. However, the presence of a non-zero $\epsilon$, leading to mixing between the singlet-like and the SM-like Higgs, allows an additional contribution from the singlet-like Higgs. Therefore, one can check for blind spots in the region where the GCE is obtained.  

At $t_\beta=1$, the up and down components of the mostly singlet like Higgs are simply related to the singlet component of the SM-like Higgs: $S_{h_{S},u} \sim S_{h_{S},d} \sim S_{h,s}/\sqrt{2}$. Including the contribution from both of these, the SI direct detection cross-section is given by:
 \begin{eqnarray}
 \sigma_{SI}&=&
 \frac{ m_p^2 m_r^2}{\pi  v^2}
   \left(\sum_{q=u,d,s} f_{Tq} + \frac{6}{27} f_{TG}\right)^2 \lambda ^2 N_{15}^4 \nonumber \\
   &&\times  \left\{\frac{S_{h,s}}{m_{h_S}^2}
   \left[ S_{h_{S},s}\left(\frac{\kappa }{\lambda }-\frac{N_{14}^2}{N_{15}^2}\right)-\sqrt{2}
   S_{h,s} \frac{N_{14} }{N_{15}}\right]-  \frac{1}{m_h^2}
\left[ S_{h,s} \left(\frac{\kappa }{\lambda
   }-\frac{N_{14}^2}{N_{15}^2}\right)+\sqrt{2}\frac{
   N_{14}}{N_{15}}\right]\right\}^2. \nonumber \\
 \end{eqnarray}
Thus one could tune away the direct detection if\begin{eqnarray}
  \frac{S_{h,s}}{m_{h_S}^2}
   \left[ \left(\frac{\kappa }{\lambda }-\frac{N_{14}^2}{N_{15}^2}\right)-\sqrt{2}
   S_{h,s} \frac{N_{14} }{N_{15}}\right] \sim \frac{\sqrt{2}}{m_h^2}
\frac{
   N_{14}}{N_{15}}.
 \end{eqnarray}
This is difficult to satisfy, however, since the light singlet mass, $m_{h_S}$ is not independent from all the other parameters. Specifically, the mixing, $S_{h,s}$ is relevant for its mass, as mentioned previously. Taking the dominant contribution to the singlet Higgs mass to scale with $\lambda$ and an upper bound on the singlet component of the SM-like Higgs, $S_{h,s} \lesssim 0.3$, to be consistent with measured Higgs properties, we find that $\lambda \lesssim 0.3$ is required in order to achieve the blind spot.  This is challenging for Higgs phenomenology at $t_\beta = 1$, since stop masses of the order of $100$~TeV or higher are required to drive the Higgs mass up to 125 GeV.  While an unpleasant region of parameter space from a UV complete point of view, this is required by the phenomenology of the GCE at $t_\beta = 1$. Therefore, similar to moderate/large $t_\beta$, even for $t_\beta=1$ we only find a viable solution where consistency with GCE, relic density, LUX constraints and SM-like Higgs phenomenology forces the allowed parameter region to be very close to resonance. We again verified our analytical results thoroughly with {\tt micrOMEGAs} and {\tt NMSSMTools}.

\subsection{Bino/Higgsino Dark Matter $(\kappa/\lambda \gg 1)$}

We now turn to the case of DM  which is an admixture of Bino and Higgsino.
We first note that in this case, the CP-even sector is effectively the MSSM Higgs sector since the singlet mass is driven up by the required large values of $\kappa/\lambda$ and is effectively decoupled. Therefore, the SM-like Higgs mass is controlled by MSSM like contributions from the squarks and there is no motivation to consider small value of $t_\beta$, which we know are problematic for obtaining a mass of 125 GeV. Hence, in this section, we will restrict ourselves to moderate/large values of $t_\beta$.

We now turn to finding the region of parameter space where Bino/Higgsino DM is viable for the GCE. 
We first consider the resonant annihilation case; as demonstrated in Fig.~\ref{FreezeoutAndGC} a large hierarchy in the couplings $g_{a\chi \chi}$ and $g_{abb}$ is needed to achieve the observed relic abundance and the GCE.  We will now show that this hierarchy  is not generally present for the Bino/Higgsino case. 

Expanding the results in Appendix~\ref{Ap:Neut} in the limit $m_\chi << \mu$, the up and down Higgsino as well as Bino parts of the neutralino can be written as
\begin{equation}
\frac{N_{13}}{N_{11}} \sim  \frac{m_Z s_W}{\mu} s_{\beta}\left(1+\frac{m_\chi}{\mu t_\beta}\right),~~~\frac{N_{14}}{N_{11}} \sim-  \frac{m_Z s_W}{\mu}c_\beta \left(1+\frac{m_\chi t_\beta}{\mu }\right),~~~N_{11} \sim \left(1 + \frac{m_Z^2 s^2_W}{\mu^2}\right)^{-1/2}.
\end{equation}
The active part of the mostly singlet pseudoscalar through which the dark matter annihilates is
\begin{equation}
\frac{P_{a,A}}{P_{a,S}} \sim - \frac{\lambda \, v}{2\mu}\left(s_{2\beta} - 6 \frac{\mu^2}{m_A^2}\frac{\kappa}{\lambda}\right) \sim 3 \kappa \frac{v \mu}{m_A^2},
\end{equation}
leading to
\begin{equation}
P_{a,S} = \left( 1+ 9\kappa^{2}\frac{v^{2}\mu^{2}}{m_{A}^{4}}\right)^{-1/2}
\end{equation}
upon normalization. The coupling of the dark matter to the lightest pseudoscalar can thus be written
\begin{eqnarray}
g_{a \chi \chi}& =& - i \sqrt{2} \left[ \lambda N_{13} N_{14}  +  m_Z s_W N_{11} (s_{\beta} N_{13} - c_\beta N_{14})\left( s_{2\beta} \frac{\lambda}{2\mu} - 3 \frac{\kappa \mu}{m_A^2}\right)\right] P_{a,S}\, , \nonumber \\
&\sim&   i 3 \sqrt{2} \kappa   \left( \frac{m_Z^2 s_W^2}{m_A^2}\right) N_{11}^2 P_{a,S} , 
\label{AchichiB/H}
\end{eqnarray}
while the coupling to $b$ quarks becomes
\begin{equation}
g_{a b b} = -i \frac{m_b s_\beta \kappa}{\sqrt{2}\mu}\left(s_\beta \frac{\lambda}{\kappa} - \frac{3 \mu^2}{c_\beta m_A^2}\right)P_{a,S} \sim  3 i \kappa \frac{m_b}{\sqrt{2}}\frac{\mu\, t_\beta}{m_A^2} P_{a,S}.
\label{AbbB/H}
\end{equation}
Note that in the above, unlike the S/H scenario, $m_A$ can be order of $|\mu|$ since we no longer have to cancel the singlet component of the SM-like Higgs. Hence the $g_{a b b}$ coupling and consequently $\sigma v$ are $t_\beta$-enhanced.
The ratio of the couplings thus becomes
\begin{equation}
\frac{g_{a \chi \chi}}{g_{a b b}} \simeq \frac{2 m_Z^2 s_W^2}{\mu m_b t_\beta},
\end{equation}
which is generically ${\cal O}(1)$, unless $\mu$ is very large.
In addition, since $\lambda$ is small, the Higgsino components in the neutralino are now much smaller than in the Singlino/Higgsino case.  As a result, the $Z$ funnel does not play an important role in setting the relic abundance. 

\begin{figure}[t]
\includegraphics[width=0.48\textwidth]{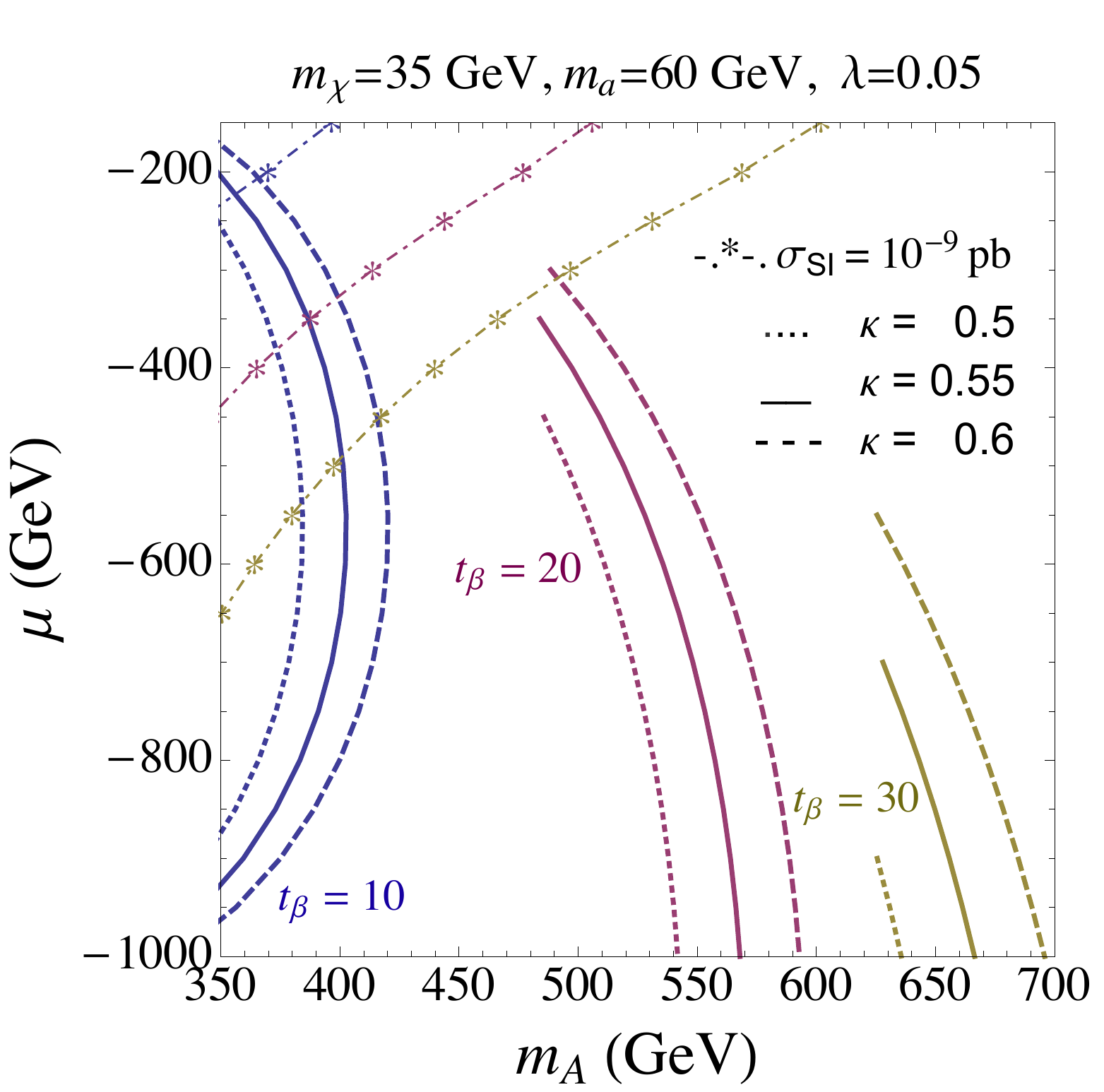} 
\includegraphics[width=0.48\textwidth]{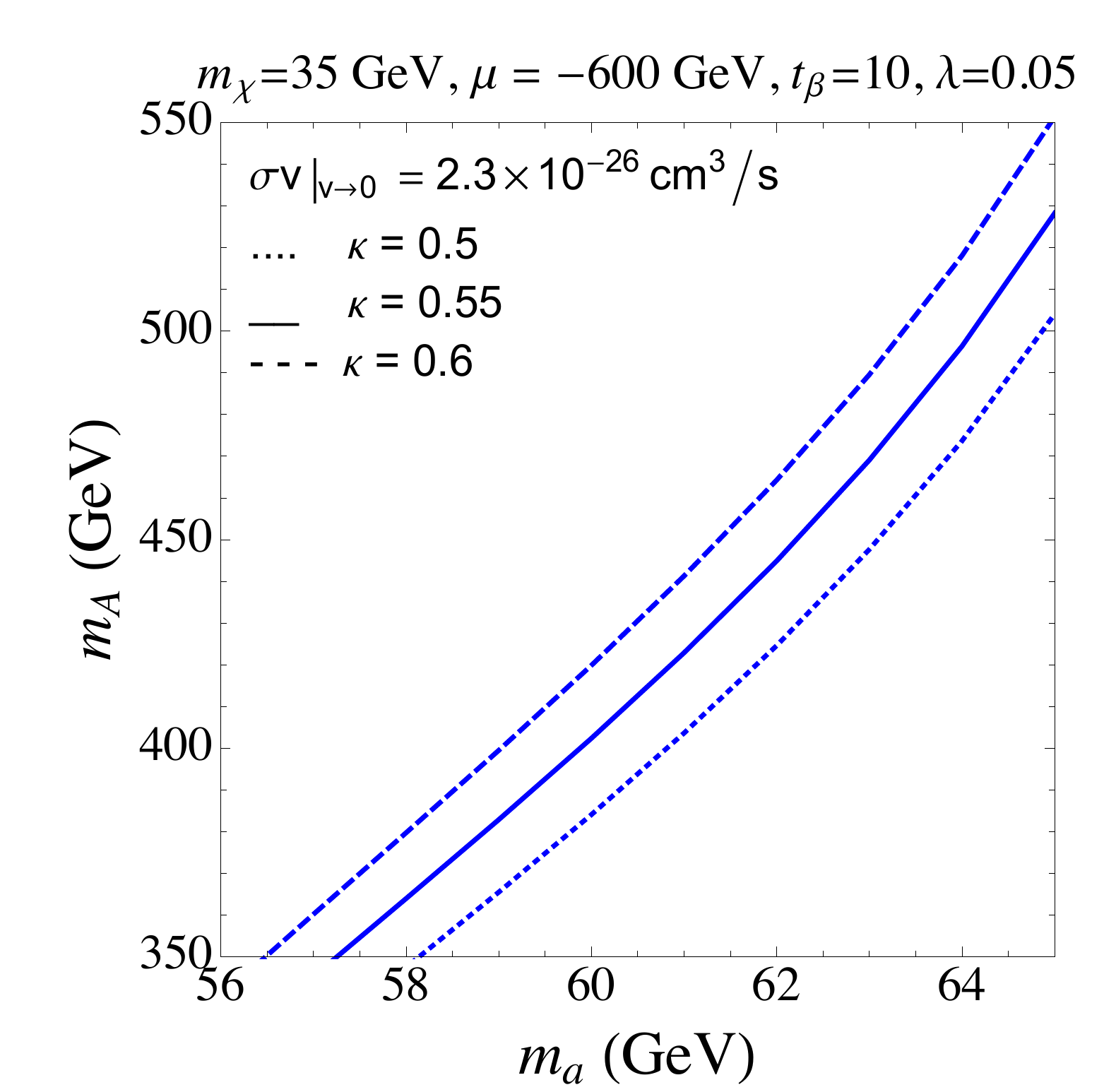} 
\caption{{\em Left:} Contours fixing the $s$-wave annihilation cross-section of Bino/Higgsino DM to $2.3 \times 10^{-26} \mbox{ cm}^3/\mbox{s}$ for various choices of $\kappa$ as a function of $\mu$ and $m_A$.  The dot-dashed lines show the constraints from requiring that the scattering cross-section in direct detection experiments be smaller than $10^{-45} \mbox{ cm}^2$ (below and to the right is allowed).  {\em Right:}  Contours fixing the $s$-wave annihilation cross-section of Bino/Higgsino DM to $2.3 \times 10^{-26} \mbox{ cm}^3/\mbox{s}$ for various choices of $\kappa$ as a function of $m_a$ and $m_A$.}
\label{2->2BinoHiggsino}
\end{figure}

We are therefore left to consider the off-resonance annihilation case, where a working solution is easily achieved for moderately large $\kappa$, and $t_\beta$ and $\mu^2/m_A^2$ not too small, as can be seen from Eqs.~\ref{AchichiB/H} and~\ref{AbbB/H}.  Utilizing these expressions, together with Eq.~\ref{eq:simpsigma}, we find the results for the GCE shown in Fig.~\ref{2->2BinoHiggsino}. In the left panel we fix the DM mass to 35 GeV and the pseudoscalar mass to 60 GeV. For a small fixed value of $\lambda=0.05$, we show the required values of $\mu$ and $m_A$ to obtain $\sigma v|_{v\rightarrow0}=2.3\times 10^{-26}$ cm$^3$/s for different values of $\kappa$ and $t_\beta$. 

Since we have fixed $m_a=60$ GeV, we are sufficiently far from resonance that the usual matching between thermal cross-section for relic density and GCE today holds. Therefore we expect that for this set of parameters, one would obtain a consistent GCE and relic density in the early universe. The right panel shows the same information but in the $m_a$ -- $m_A$ plane with a fixed value of $\mu=-600$ GeV and $t_\beta$=10. The hard cut-off for each value of $t_\beta$ in the left panel for $m_A$ is due to a naive implementation of the  LHC $H/A\rightarrow \tau^+ \tau^-$ bounds~\cite{CMS-PAS-HIG-13-021}, assuming that  both $m_H$ and $m_{a_2}$ are approximately given by $m_A$. However, note that in this scenario, there can be significant mixing between the two pseudoscalars, thereby changing the correlation of $m_{a_2}$ with $m_A$. On one hand this could lead to a weakening of these bounds for a given $m_A$, but on the other hand this could strengthen them due to the presence of a large active component in $m_a$. We will discuss this and other relevant constraints due to Higgs phenomenology in more detail when we analyze our full numerical results obtained from {\tt micrOMEGAs} and {\tt NMSSMTools}.

The parameter region under consideration is also easily made compatible with LUX limits.
The scattering cross-section for B/H DM through $h$ and $H$ is given by 
\begin{eqnarray}
\sigma_{SI} &\simeq&  \frac{ m_Z^2 s_W^2 m_p^2 m_r^2
}{\pi  v^4
  }  N_{11}^4  \left[\frac{\left(\frac{F_u}{t_{\beta }}-F_d
   t_{\beta }\right) }{m_H^2}\left(\frac{N_{14} }{N_{11}}c_{\beta
   }+\frac{N_{13}
   }{N_{11}}s_{\beta
   }-\frac{\lambda v}{m_Z s_W}\frac{N_{13}}{N_{11}}\frac{   N_{14} 
  }{N_{11} } S_{H,s}\right) \right. \nonumber \\
  &&\qquad \qquad \left.  -\frac{\left(F_d+F_u\right)}{m_h^2} \left(\frac{N_{13} }{N_{11}}c_{\beta
   }-\frac{N_{14} }{N_{11}}s_{\beta
   }\right)\right]^2,  \nonumber \\ 
   \\
   &\sim& \frac{m_p^2 m_r^2
  }{\pi  v^4} \frac{m_Z^4 s_W^4}{\mu^4}N_{11}^4  \left[\frac{\left(F_d+F_u\right) }{m_h^2}\left(m_{\chi
   }+\frac{2 \mu }{t_{\beta
   }}\right)+\frac{
   F_d}{m_H^2}\mu t_{\beta }\right]^2,
\end{eqnarray}
where in the second line we have used the large $t_\beta$
approximations and kept only the leading Higgsino contributions. This
is exactly equivalent to the MSSM direct detection cross-section at
large $t_\beta$. In this case, opposite to the S/H case, negative
$\mu$ tends to suppress the direct detection
cross-section~\cite{Feng:2010ef, Huang:2014xua}.  This suppression
occurs both via the $h\chi\chi$ coupling and the interplay of $h$- and
$H$-mediated annihilation diagrams, allowing for significant freedom
to evade direct detection constraints.  The relative size of the $M_1$
and $\mu$ terms also suppressed the spin-dependent direct detection
cross-section, which is at least two orders of magitude beyond current
bounds~\cite{Aprile:2013doa}.  In the left panel of
Fig.~\ref{2->2BinoHiggsino}, in addition to the required values for
GCE, we show the contours where $\sigma_{SI}=10^{-9}$ pb for different
values of $t_\beta$ in the $\mu-m_A$ plane. To the right of these
contours, the direct detection cross-section therefore does not
provide a relevant constraint.

\begin{figure}[t]
\includegraphics[width=0.8\textwidth]{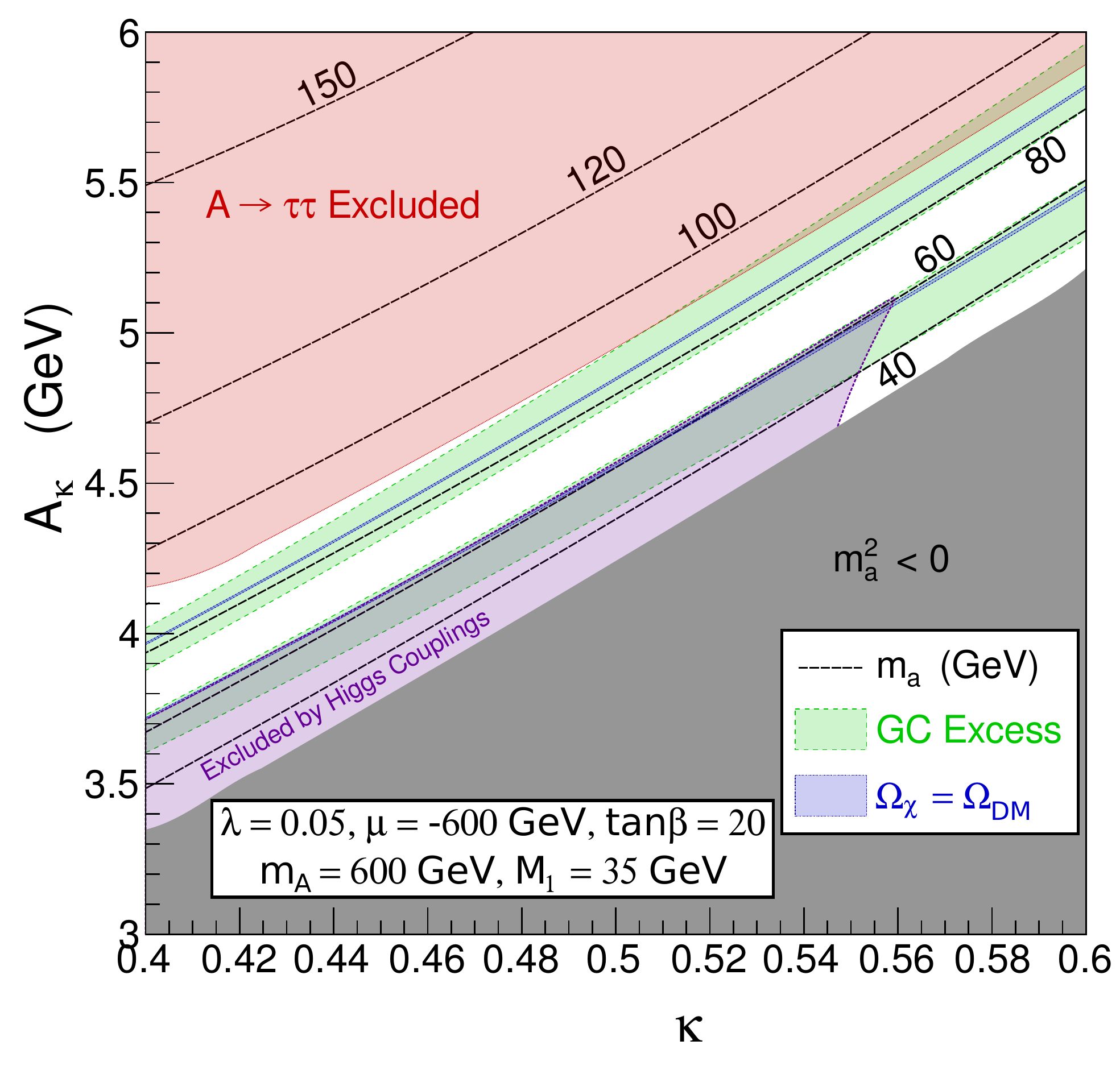}
\caption{Results of numerical scan with $M_1 =$ 35 GeV
$\mu =$ -600 GeV,
$m_A =$ 600 GeV,
$t_\beta =$ 20,
$\lambda =$ 0.05, 122 GeV $ < m_h < $ 128 GeV, GCE: $0.5 <  10^{26}$ cm$^3$/s$ \times \sigma v|_{v\rightarrow 0} <$  4, 0.1118 $ < \Omega h^2 <$ 0.128.  The green and blue regions show the parameter space consistent with the GCE and observed relic density.  The black dashed lines show the contours for constant pseudoscalar mass, $m_{a}$. Notice that as $\kappa$ increases, the regions with the correct relic abundance pull slightly away from resonance, as expected from our analytical results. The red region is excluded by $A\rightarrow \tau^{+}\tau^{-}$ searches at the LHC, while the purple exclusion comes from modification of SM Higgs rates due to the presence of the open $h\rightarrow a\, a$ channel.}
\label{2->2BHMO}
\end{figure}

The viable region for Bino/Higgsino DM is summarized in
Fig.~\ref{2->2BHMO} where we present the results of a full numerical
scan using {\tt micrOMEGAs} and {\tt NMSSMTools}.  The
parameter space is set by
$(\lambda,\,\kappa,\,A_{\kappa},\,\mu,\,t_{\beta},\,M_{1},\,m_{A})$. For
each point in the scan, without loss of generality, we have fixed $M_1
= 35\,{\rm GeV}$, producing the value $m_\chi \approx 35~\gev$
favored by the GCE. We also fix $\lambda = 0.05$, which, as can be
seen from the expressions for the pseudoscalar couplings, does not
affecting the phenomenology if sufficiently small. Furthermore
$t_{\beta}$ was fixed to 20 and $\mu$ and $m_{A}$ were fixed to
$-600\,{\rm GeV}$ and $600\,{\rm GeV}$ respectively, sufficiently
heavy to evade direct detection LHC bounds. Therefore we are left with two parameters, $\kappa$
and $A_{\kappa}$, taken as the axes of Fig.~\ref{2->2BHMO}. They
control the couplings and the lightest pseudoscalar mass respectively
as discussed in Appendix~\ref{app:BH}. We further fix all other soft
masses to 1 TeV, with the exception of the stop sector, where we fix
$A_t=\sqrt{6}\, m_{Q_3}$, and $m_{Q_3}=m_{u_3}=7.5$ TeV resulting
in a SM-like Higgs mass in the range 122-128 GeV across the plane. 

As one can see, the GCE allowed regions (green) and the correct relic density (blue) overlap along two stripes in the $(\kappa,\,A_{\kappa})$ plane and are close to the  regions where $2m_{\chi}$ and $m_{a}$ differ by about $20\%$, consistent with off-resonance conditions discussed in Sect.~\ref{Sec:Simplified}.

As can be seen from Eq.~\ref{eq:BHmix}, and mentioned previously, the
lightest pseudoscalar has a non-negligible active component, up to
$50\%$ in the region of interest, rendering it quite
MSSM-like. Consequently, this state is constrained by collider
results. If $m_{a}$ is sufficiently light, decays of the SM-like Higgs
into a pair of pseudoscalars are open and significantly modify
Higgs coupling measurements. The overall contribution depends on the
$haa$ coupling, which is controlled by $\lambda A_\lambda$ but has
sub-leading contributions due to $\kappa$ and
$A_{\kappa}$~\cite{Ellwanger:2009dp}. This excludes much of the lower
branch consistent with the GCE where $m_{a}\lesssim 60\,{\rm GeV}$,
though the bound weakens for $\kappa \gtrsim 0.55$ because the $haa$
coupling becomes sufficiently small.

$H/A\rightarrow \tau^{+}\tau^{-}$ bounds~\cite{CMS-PAS-HIG-13-021} are also
significant, excluding a portion of the upper branch consistent with
the GCE for $m_{a}>90\, {\rm GeV}$.  Both the pseudoscalars and the heavy scalar are relevant
for this constraint.  The heavier pseudoscalar has a mass of
$500-700\gev$ throughout the plane, which is sufficiently large that
it evades $H/A\rightarrow \tau^{+}\tau^{-}$ bounds even without a
singlet component to further suppress the production cross-section.
This is remarkable because the mass of the heavier MSSM scalar and
charged Higgs is $450$ GeV throughout the plane~\footnote{The
  significant discrepancy between $m_A$ and the heavier MSSM-like
  scalar mass is due to radiative corrections.}, for which
$H/A\rightarrow \tau^{+}\tau^{-}$ bounds require $t_\beta < 18.8$.
Thus the bounds are evaded partially because the NMSSM allows the
scalar and pseudoscalar mass and mixing structures to be decoupled.

Meanwhile, LHC $H/A\rightarrow \tau^{+}\tau^{-}$ limits on a $90\gev$
MSSM-like pseudoscalar require $t_{\beta} < 7.19$. The production cross-section
is suppressed somewhat due to the singlet component of the lightest
pseudoscalar and the fact that no CP-even scalar accompanies it, but
the MSSM-like component is still large enough to result in a strong
bound. The pseudoscalar
production cross section scales with $t_{\beta}^{2}$, so that the
entire region with $m_{a}\gtrsim 90\,{\rm GeV}$ in Fig.~\ref{2->2BHMO}
is excluded since we set $t_{\beta}=20$.    Increasing $m_{A}$ would reduce the active admixture of the
lightest pseudoscalar, but the GCE requires keeping the ratio
$t_{\beta}/m_{A}$ approximately constant so that this LHC exclusion is quite
robust\footnote{Reducing both $m_{A}$ and $t_{\beta}$ may allow one to evade the
  limits, but it would create more tension in obtaining a SM-like Higgs
  with a mass of $125\,{\rm GeV}$, as is well known for the
  MSSM.}. This exclusion is shown as the red region in
Fig.~\ref{2->2BHMO}.  Below $m_{a}\sim 90\,{\rm GeV}$ there are no
published limits from the LHC and the LEP limits from
$e^{+}e^{-}\rightarrow hA$~\cite{Schael:2006cr} are too weak to
exclude the region.

In addition, there are flavor constraints coming from $B_s \rightarrow \mu^{+}\mu^{-}$. In general a suppression of ${\cal O}(10)$ at the level of the amplitude may be required at large $t_{\beta}$. This is however easy to achieve. From a low energy point of view, there are various ways to ensure the consistency of the models with the measured value~\cite{CMSandLHCbCollaborations:2013pla}, even without relaxing the assumption of minimal flavor violation (MFV)~\cite{Altmannshofer:2012ks}. Cancellations can occur between the wino- and gluino-mediated contributions against the Higgsino contribution, depending on the sign of $A_{t}$. Moreover the wino and gluino contributions, which depend more strongly on $t_{\beta}$, can be further suppressed by requiring alignment of the squark mass matrices in the down sector. All these various options to ensure consistency with the $B_{s}\rightarrow \mu^{+}\mu^{-}$ measurement may require additional model building efforts within a UV-complete model addressing the SUSY flavor problem, which are beyond  the scope of this paper. Therefore we will not discuss flavor constraints further.

We would like to stress that 
the Bino/Higgsino case realizes the
original purpose of going to the NMSSM to relax the relations between
the lightest pseudoscalar mass and the charged/CP-even Higgs masses that were
obstructing a viable GCE model within the MSSM.  The direct detection
cross section and two lighter neutral
scalars are very MSSM-like, but the parameter regions consistent with
the GCE and thermal relic density are possible because the light
pseudoscalar has a sizable component of both the MSSM and the singlet
pseudoscalars.  The viable B/H region with a light and fairly active
pseudoscalar \emph{without} either a light charged Higgs or a CP-even Non-SM-like Higgs presents interesting
challenges for the LHC Run II: Direct searches for a
pseudoscalar with mass $\sim 60-90\,{\rm GeV}$ and production cross
section suppressed by a factor of ${\cal O}({\rm few})$ compared to
the MSSM, may still reveal a signal in a region left open by LEP due
to its kinematic limits. This region, normally deemed excluded in the
conventional MSSM $(m_{A},\, t_\beta)$ plane, requires dedicated
detailed studies of the LHC signatures, which are beyond the scope of
this paper and are left for the future.

\section{Summary and Conclusions}
\label{Sec:Conc}

We have examined models to explain the GCE with thermal relic DM, focusing on the $\chi\chi\rightarrow b \bar b$ annihilation in supersymmetric models. We have found that while the MSSM fails to accommodate the GCE due to SM-like Higgs precision,  LHC $H^{\pm}\rightarrow \tau \nu$ searches and the mass relations between the pseudoscalar and charged/CP-even Higgs bosons, viable regions can be found in the $Z_3$ NMSSM. Both Singlino/Higgsino and Bino/Higgsino DM can explain the excess. 

In the case of Singlino/Higgsino DM, the mostly singlet pseudoscalar
is light, and there is an accompanying light CP-even.  The parameters
of the mass matrix must then be tuned to ensure that the SM-like Higgs
does not pick up a large singlet component, and to keep the other
light CP-even state mostly singlet.  In addition, $2 m_\chi$ must be within a few percent of the light pseudoscalar mass, $m_a$, to ensure compatibility with both the GCE and the observed relic abundance. 

For the Bino/Higgsino case, the situation is much less tuned: an
$\mathcal{O}(1)$ value for $\kappa$, moderate $t_\beta$, and negative
$\mu$ of at least several hundred GeV allow one to achieve the
observed GCE and relic abundance well away from any resonant region
without inducing too large of a nucleon scattering cross-section. This
closely parallels the would-be MSSM solution, which works here given
the extra freedom provided in the NMSSM for decoupling the charged/CP-even and
the lightest pseudoscalar Higgs masses. Given the large MSSM-like
fraction of the lightest pseudoscalar, up to $50\%$, this region
provides interesting LHC Higgs phenomenology worth further study. In
particular, extending LHC pseudoscalar Higgs searches below the
current mass threshold of $90\,{\rm GeV}$ would probe a large fraction
of the parameter space relevant for the GCE.

Because of the peculiar requirements of these $2 \rightarrow 2$
models, one may advocate for looking beyond the $2 \rightarrow 2$
annihilation models into $2 \rightarrow 4$ annihilation.  This has
already been considered for the general NMSSM, where annihilation
occurred to decoupled singlet pseudoscalars in a mostly decoupled hidden sector~\cite{Berlin:2014pya}. Within the $Z_3$ NMSSM the needed spectrum is difficult to achieve, because the parameters needed to obtain a large enough annihilation rate tend to induce a problematic Higgs sector; we leave examination of these models for future work.  In conclusion, while achieving the GCE excess via the MSSM is very difficult, simple viable models exist within the NMSSM. 

 \begin {acknowledgements}
We thank Tracy Slatyer for collaboration in the early stages of this work.  NRS also thanks C. Wagner and T. Liu for useful discussion. This work was supported in part by the National Science Foundation under Grant No. PHYS-1066293 and the hospitality of the Aspen Center for Physics. Work at KITP is supported by the National Science Foundation under Grant No.~NSF PHY11-25915. NRS is supported by the DOE grant No.~DE-SC0007859 and by the Michigan Center for Theoretical Physics.  KZ is supported by the DOE and by NSF CAREER award PHY 1049896.   DS is supported in part by U.S. Department of Energy grant DEÐFG02Ð92ER40701 and by the Gordon and Betty Moore Foundation through Grant No. 776 to the Caltech Moore Center for Theoretical Cosmology and Physics. CC is supported by a DOE Early Career Award DE-SC0010255.
\end{acknowledgements}

\appendix

\section{General NMSSM}

We will follow notations and conventions consistent with
Ref.~\cite{Ellwanger:2009dp} where the full super potential and all possible soft breaking terms for the general NMSSM are detailed. We present here all the relevant mass matrices and mixing angles
in the general NMSSM before reducing to the $Z_3$ case. Where relevant we will denote the matrices in the $Z_3$ NMSSM with a subscript. However, we drop this subscript in the main text since only the $Z_3$ NMSSM is discussed in detail there. Throughout,

\begin{eqnarray}
v=\sqrt{v_u^2+v_d^2}&=&174~ \textrm{GeV}\\
t_\beta\equiv\tan\beta &=& \frac{v_u}{v_d}\\
\mu \equiv \mu_{eff} & = & \mu + \lambda s \\
B \equiv B_{eff} & = & A_\lambda + \kappa s \\
\hat{m}_3^2 & = & m_3^2 + \lambda \mu' s
\end{eqnarray}
where $v_u$, $v_d$ and $s$ are the vacuum expectation values of $H_u$, $H_d$ and $S$ respectively.

\section{Neutralino Masses and Mixings}
\label{Ap:Neut}

The neutralino mass matrix is:
\begin{equation}
{\cal M} = \left(
\begin{array}{ccccc}
M_1 & 0 & -\frac{g_1 v_d}{\sqrt{2}} & \frac{g_1 v_u}{\sqrt{2}} & 0 \\
  & M_2 & \frac{g_2 v_d}{\sqrt{2}} & - \frac{g_2 v_u}{\sqrt{2}} &0 \\
& & 0 & -\mu_{eff} & -\lambda v_u \\
& & & 0 & -\lambda v_d\\
& & & & 2 \kappa s + \mu'
\end{array}
\right). \label{eq:MN}
\end{equation}
The lightest mass eigenstate of the neutralino is defined in terms of its components as:
\begin{equation}
\chi = N_{11} \tilde{B}+N_{12} \tilde{W}+N_{13} \tilde{H}_d+N_{14}\tilde{H}_u+N_{15}\tilde{S}.
\end{equation}

The characteristic equation for the neutralinos is, for $m_\chi \neq |\mu|$,
\begin{eqnarray}
0 & = & - \lambda^2 v^2 \left\{ \left( m_\chi - M_1 \right) \left( m_\chi - M_2
  \right) \left( m_\chi - 2 \mu \frac{v_u v_d}{v^2} \right)
  - \frac{1}{2} g_2^2 v^2 \left[ m_\chi - M_1 + \left( m_\chi - M_2 \right) \tan^2
  \theta_W \right] \right\}  \nonumber\\
\nonumber && + \left( m_\chi - 2 \kappa s - \mu' \right) \left\{ \left( m_\chi -
  M_1 \right) \left( m_\chi - M_2 \right) \left( m_\chi^2 - \mu^2
  \right)  \textcolor{white}{\frac{1}{2}}\right. \\
\nonumber && \left. \hspace*{3cm} - \frac{1}{2} g_2^2 v^2 \left( m_\chi + 2
\mu \frac{v_u v_d}{v^2} \right) \left[ m_\chi - M_1 + \left(
  m_\chi - M_2 \right) \tan^2 \theta_W \right] \right\}.\nonumber \\
\end{eqnarray}

If we decouple the wino, the above reduces to:
\begin{eqnarray}
0 & = & - \lambda^2 v^2 \left[ \left( m_\chi - M_1 \right) \left( m_\chi - 2
  \mu \frac{v_u v_d}{v^2} \right) - \frac{1}{2} g_2^2 v^2
  \tan^2 \theta_W \right] \\
\nonumber && + \left( m_\chi - 2 \kappa s - \mu' \right) \left[ \left( m_\chi -
  M_1 \right) \left( m_\chi^2 - \mu^2 \right) - \frac{1}{2}
  g_2^2 v^2 \left( m_\chi + 2 \mu \frac{v_u v_d}{v^2} \right)
  \tan^2 \theta_W \right]
\end{eqnarray}

We will concentrate on two limiting cases for the composition of the neutralino: Singlino/Higgsino and Bino/Higgsino.

\subsection{Singlino/Higgsino}


\begin{itemize}
\item{
 General NMSSM:	$2\frac{\kappa \mu}{\lambda} +\mu' << \mu <<  M_1$,   }
 
 	Using the characteristic polynomial, and also decoupling the Bino, we can trade $\mu'$ for the mass eigenvalue, $m_\chi$:
	\begin{equation}
	\mu'  =   m_\chi - 2 \kappa \frac{\mu}{\lambda}
  - \frac{\lambda^2 v^2 \left( m_\chi - \mu s_{2\beta} \right)}{m_\chi^2 - \mu^2}. 
	\end{equation}

 \item{$Z_3$ NMSSM:	$2\frac{\kappa}{\lambda}  << 1$,  }
  
  $\mu'=0$ in the $Z_3$ NMSSM, therefore, we can instead re-write $\kappa$ in terms of the mass eigenvalue $m_\chi$:

 \begin{eqnarray}
\kappa & = & \frac{\lambda}{2 \mu} \left[ m_\chi 
  - \frac{\lambda^2 v^2 \left( m_\chi - \mu s_{2\beta}
    \right)}{m_\chi^2 - \mu^2} \right]. \label{eq:kappa}
\end{eqnarray}
 \end{itemize}
 
In both cases:
 \begin{eqnarray}
\frac{N_{13}}{N_{15}}&=& \frac{\lambda  v }{\mu ^2-m_{\chi }^2}c_\beta \left(t_\beta m_{\chi }-\mu \right) \sim  \frac{\lambda  v }{\mu}c_\beta \left(t_\beta \frac{m_{\chi }}{\mu}-1\right),  \\
\frac{N_{14}}{N_{15}}&=&\frac{-\lambda  v }{\mu ^2-m_{\chi }^2}s_\beta \left(\mu -\frac{m_{\chi }}{t_\beta}\right) \sim -\frac{\lambda  v }{\mu} s_\beta, \\
N_{15} &=&\left(1+ \frac{N^2_{13}}{N^2_{15}}+\frac{N^2_{14}}{N^2_{15}}\right)^{-1/2}.
\end{eqnarray}

\subsection{Bino/Higgsino} \label{app:BH}
\begin{itemize}
\item{    General NMSSM:	$2\frac{\kappa \mu}{\lambda} +\mu' >> \mu >>  M_1$,}
	
\item{ $Z_3$ NMSSM:	$2\frac{\kappa}{\lambda}  >> 1$,  }

\end{itemize}
In both cases, we decouple the Singlino, and therefore $M_1$ can be re-written in terms of the mass eigenvalue, $m_\chi$:

\begin{equation}
M_1=m_{\chi }+\frac{m_Z^2 s_W^2 \left(\mu  s_{2\beta}+m_{\chi }\right)}{\mu^2-m_{\chi }^2}.
\end{equation}

The components are then given by:

\begin{eqnarray}
N_{11}&=&\left(1+\frac{N^2_{13}}{N^2_{11}}+\frac{N^2_{14}}{N^2_{11}}\right)^{-1/2},\\
\frac{N_{13}}{N_{11}}&=&\frac{m_Z s_W s_\beta }{\mu ^2-m_{\chi }^2}  \left(\mu  +\frac{m_{\chi }}{t_\beta}\right) \sim \frac{m_Z s_W  }{\mu} s_{\beta},\\
\frac{N_{14}}{N_{11}}&=&-\frac{m_Z s_W c_\beta }{\mu^2-m_{\chi }^2} \left(\mu +t_\beta m_{\chi }\right) \sim -\frac{m_Z s_W}{\mu} c_{\beta}\left(1+t_{\beta}\frac{m_\chi}{\mu}\right).
\end{eqnarray}

\section{CP-Odd Mass Matrix}
The general CP-odd mass matrix in the $(A,S)$ ``interaction'' basis is given by
\begin{eqnarray}
\mathcal{M}^2_P & = & \left(
\begin{array}{cc}
\frac{2 (\mu B + \hat{m}_3^2)}{s_{2\beta}}
& \lambda (A_\lambda -2 \frac{\kappa \mu}{\lambda} -\mu') v
\\
&
-\frac{3 \kappa  \mu  A_{\kappa }}{\lambda }+\frac{\lambda ^2 v^2 }{2\mu }s_{2\beta} 
   \left(B+\frac{3 \kappa  \mu }{\lambda }+\mu '\right)-\xi _F \left(4 \kappa +\frac{\lambda  \mu
   '}{\mu }\right)-\frac{\kappa  \mu  \mu '}{\lambda }-2 m_s'^2  
\\
\end{array} 
\right).\nonumber\\
\end{eqnarray}

Defining $m_A^2$ as the (1,1) element of the above matrix: $m_A^2  = 2 \left( \mu B
+ \hat{m}_3^2 \right)  \csc 2\beta$, 

\begin{eqnarray}
\mathcal{M}^2_P & = & \left(
\begin{array}{cc}
m_A^2
& \lambda (A_\lambda -2 \frac{\kappa \mu}{\lambda} -\mu') v
\\
&
-\frac{3 \kappa  \mu  A_{\kappa }}{\lambda }+\frac{\lambda ^2 v^2}{2\mu } s_{2\beta}
   \left(A_{\lambda }+\frac{4 \kappa  \mu }{\lambda }+\mu '\right)-\xi _F \left(4 \kappa
   +\frac{\lambda  \mu '}{\mu }\right)-\frac{\kappa  \mu  \mu '}{\lambda }-2 m_s'^2\\
\end{array} 
\right).\nonumber\\
\end{eqnarray}

Further using the characteristic polynomial for the above, we can redefine $m_s'^2$ in terms of the lighter mass eigenvalue, $m_a$, and all the other parameters:

\begin{eqnarray}
{m'_S}^2 & = & - \frac{\lambda^2 v^2 \left( A_\lambda - 2
  \frac{\kappa \mu}{\lambda} \right)^2}{2 \left( m_A^2
  - m_a^2 \right)} - \frac{m_a^2}{2} + \left( A_\lambda + 4
\frac{\kappa \mu}{\lambda} \right) \frac{\lambda^2
  v^2}{4\mu} s_{2\beta} - \frac{3}{2} \kappa A_\kappa
s \nonumber  \\
&&- \frac{1}{2} \kappa \mu' s + \frac{1}{2} \xi_F \left( 4 \kappa + \frac{\mu'}{s} \right) + \frac{\xi_S}{2s}.
\end{eqnarray}

In the absence of the singlet, $m_A$ would be the usual MSSM parameter controlling the CP-odd Higgs mass as well as the CP-even non-standard Higgs. 

In the limit that $m_A^2 >> m_a^2$, 
the components of  $a$ are given by:
\begin{eqnarray}
\frac{P_{a,A}}{P_{a,S}} &\approx& -\frac{ \lambda v}{ m_A^2}(A_\lambda -2 \frac{ \kappa \mu}{\lambda} -\mu')\, , \\
P_{a,S} &=&\left(1+\frac{P_{a,A}^2}{P_{a,S}^2}\right)^{-1/2},
\end{eqnarray}
where $P_{a,A}$ is the active component and $P_{a,S}$ is the singlet component of the light CP-odd Higgs.

If we now take $A_\lambda$ such that we minimize the mixing with the SM-like Higgs, 
\begin{equation}
\frac{P_{a,A}}{P_{a,S}}\approx -\frac{2\lambda  v}{m_A^2} \left(\frac{\mu}{s_{2\beta}}- \left(2 \frac{\kappa  \mu }{\lambda }+\mu
   '\right)+\frac{\epsilon}{2} \right).
\end{equation}

In terms of the components given above, the relevant couplings of the light CP-odd Higgs are:
\begin{eqnarray}
y_{abb}&=&\frac{i m_b  t_{\beta }}{\sqrt{2} v}P_{a,A} \\
y_{a\chi\chi}&=& i  \left\{ \left[\left(N_{14} c_{\beta }-N_{13} s_{\beta
   }\right) \left( g_1N_{11}-g_2 N_{12} \right) + \sqrt{2}\lambda N_{15}
   \left(N_{13} c_{\beta }+N_{14} s_{\beta }\right)\right] P_{a,A}\right. \nonumber \\
   &&  \left. +\sqrt{2}\left(\lambda  N_{13} N_{14}
    - \kappa N_{15}^2\right)
 P_{a,S}  \right\}. \nonumber \\
\end{eqnarray}

In the $Z_3$ NMSSM, $m_3=0$ and $A_\lambda$ is no longer a free parameter, but is related to $m_A$ via Eq.~\ref{eq:mAAlambda}. The mass matrix reduces to: 

\begin{eqnarray}
\mathcal{M}^2_{P_{Z_3}} & = & \left(
\begin{array}{cc}
m_A^2
&\lambda  v \left(\frac{m_A^2 }{2\mu }s_{2\beta} -\frac{3 \kappa  \mu }{\lambda }\right)
\\
&
\lambda ^2 v^2 s_{2\beta}  \left(\frac{m_A^2}{4\mu ^2} s_{2\beta}+\frac{3 \kappa }{2\lambda }\right)-\frac{3
   \kappa  A_{\kappa } \mu }{\lambda }\\
\end{array} 
\right).\nonumber\\
\end{eqnarray}
Now, we can use the characteristic polynomial for the CP-odd mass matrix to re-write $A_\kappa$ in terms  $m_a$:
\begin{equation}
A_\kappa  =     -\frac{\lambda }{3 \kappa  \mu } \left[m_a^2 -\frac{\lambda ^2 v^2 s_{2\beta }
   }{2\mu }\left(\frac{m_A^2 s_{2\beta }}{2\mu }+\frac{3 \kappa  \mu }{\lambda
   }\right) -\frac{\lambda ^2 v^2 }{m_a^2-m_A^2}\left(\frac{m_A^2 
   s_{2\beta }}{2\mu }-\frac{3 \kappa  \mu }{\lambda
   }\right)^2 \right].
\label{eq:akappa}
\end{equation}

After further requiring minimal mixing of the SM-like CP-even scalar with the singlet, the active component of $a$ is given by:

\begin{eqnarray}
\frac{P_{a,A}}{P_{a,S}}&\approx& -\frac{\lambda  v}{m_A^2} \left( \frac{m_A^2}{2\mu} s_{2\beta}-3 \frac{\kappa  \mu }{\lambda } \right)\nonumber \\
 &\sim&  - \frac{\lambda  v}{m_A^2} \left( \frac{2\mu}{s_{2\beta}}-4 \frac{\kappa  \mu }{\lambda }+\epsilon\frac{s_{2\beta}}{2\mu} \right). \label{eq:BHmix} 
\end{eqnarray}

\section{CP-Even Mass Matrix}
\label{sec:CPE}

In the basis $\left( H_{d}, H_{u}, S \right)$ the general mass matrix for
the CP-even scalars is 

{\tiny
\begin{eqnarray}
\mathcal{M}^2_R & = & \left(
\begin{array}{ccc}
g^2 v^2 c^2_{\beta} + \left( \mu B +
\hat{m}_3^2 \right) t_{\beta} & \left(\lambda^2 - \frac{1}{2} g^2
\right) v^2 s_{2\beta} - \mu B -
\hat{m}_3^2 & \lambda \left( 2 \mu v c_{\beta} -
\left(B + \kappa s + \mu'\right) v s_{\beta} \right) \\
\left(\lambda^2 - \frac{1}{2} g^2 \right) v^2 s_{2\beta} -
\mu B - \hat{m}_3^2 & g^2 v^2 \sin^2
\beta + \left( \mu B + \hat{m}_3^2
\right) \cot\beta & \lambda \left( 2 \mu v s_{\beta} -
\left(B + \kappa s + \mu'\right) v c_{\beta} \right) \\
\lambda \left( 2 \mu v c_{\beta} -
\left(B + \kappa s + \mu'\right) v s_{\beta} \right) &
\lambda \left( 2 \mu v s_{\beta} -
\left(B + \kappa s + \mu'\right) v c_{\beta} \right) &
\frac{1}{2} \lambda \left(A_\lambda + \mu' \right) \frac{v^2}{s} s_{2\beta} + \kappa s \left( A_\kappa + 4 \kappa s + 3 \mu' \right) -
\frac{\xi_S + \xi_F \mu'}{s}
\end{array}
 \right)\nonumber \\
\end{eqnarray}
}


Rotating the upper $2\times 2$ matrix by the angle $\beta$ and replacing $M_A^2  = 2 \left( \mu B
+ \hat{m}_3^2 \right)/ s_{2\beta}$ gives now the mass matrix in the (H,h,S) basis: 

{\tiny
\begin{eqnarray}
\mathcal{M}^2_h & = & \left(
\begin{array}{ccc}
m_A^2+s^2_{2\beta} \left(m_Z^2-\lambda ^2 v^2\right)
& 
s_{2\beta} c_{2\beta} \left(m_Z^2-\lambda ^2 v^2\right)
&
-\lambda v c_{2\beta} \left(A_{\lambda }+\frac{2 \kappa  \mu }{\lambda }+\mu '\right)
\\
s_{2\beta} c_{2\beta} \left(m_Z^2-\lambda ^2 v^2\right)
  & 
c^2_{2\beta} m_Z^2+\lambda ^2 v^2 s^2_{2\beta}
& 
2 \lambda  v \left(\mu -s_{\beta}c_{\beta} \left(A_{\lambda }+\frac{2 \kappa  \mu }{\lambda }+\mu '\right)\right)
\\
-\lambda v c_{2\beta} \left(A_{\lambda }+\frac{2 \kappa  \mu }{\lambda }+\mu '\right)
& 
2 \lambda  v \left(\mu -s_{\beta}c_{\beta} \left(A_{\lambda }+\frac{2 \kappa  \mu }{\lambda }+\mu '\right)\right)
& 
\frac{1}{2}  \left(A_\lambda +
\mu' \right) \frac{\lambda^2 v^2}{\mu} s_{2\beta} + \frac{\kappa  \mu}{\lambda} \left( A_\kappa + 4
\frac{\kappa  \mu}{\lambda} + 3 \mu' \right) - \frac{\lambda}{\mu}(\xi_S + \xi_F \mu')
\end{array}
 \right)\nonumber\\
\end{eqnarray}
}
Note that in the absence of the singlet, the upper $(2\times2)$ matrix is the MSSM Higgs mass matrix and it is clear that these fields would acquire expectation values according to: $<h>=v$, and $<H>=0$, clarifying our notation. 

If we further set $A_\lambda$ such that the mixing of the singlet with the SM-like Higgs is $\epsilon$, the off-diagonal terms mixing with the singlet reduce to:

\begin{eqnarray}
A_\lambda &=& \frac{2 \mu}{s_{2\beta}}-\frac{2 \kappa  \mu }{\lambda }-\mu '+\epsilon\, ,\\
\mathcal{M}^2_h(1,3)&=& \lambda  v  c_{2\beta} \left(\frac{2\mu}{s_{2\beta}}+\epsilon \right)\, ,\\
\mathcal{M}^2_h(2,3)&=&-\lambda  v \epsilon  s_{2\beta}\,.
\end{eqnarray}

The mass eigenstates are defined by in terms of the components $S_{i,j}$ where $i=\{H,h,h_{s}\}$ 
and $j=\{u,d,s\}$:
\begin{eqnarray}
H&=& S_{H,d} H_d + S_{H,u} H_u -S_{H,s} S\, ,\\
h&=& S_{h,d} H_d + S_{h,u} H_u -S_{h,s} S\, ,\\
h_S&=& S_{h_{S},d} H_d + S_{h_{S},u} H_u +S\, .
\end{eqnarray}
%
We will assume that the non-singlet-like non-standard Higgs, $H$, is decoupled from the mostly SM-like Higgs, $h$. This implies that the up and down components of $H$ and $h$ are given as in the usual MSSM decoupling limit by
\begin{eqnarray}
S_{H,d}=S_{h,u}&\equiv& c_\alpha=s_\beta,\\
S_{H,u}= - S_{h,d} &\equiv& -s_\alpha = c_\beta.
\end{eqnarray}
Further, we will be always interested in the case when the singlet is mostly decoupled from the other two CP-even Higgses. In such a case, the singlet components of the standard and the non-standard Higgses $S_{h,s}$ and $S_{H,s}$, are related to the up and down components of the mostly singlet CP-even Higgs, $S_{h_{S},u}$ and $S_{h_{S},d}$ :
\begin{eqnarray}
S_{h_{S},d}&=& S_{h,s} c_\beta + S_{H,s} s_\beta, \\
S_{h_{S},u}&=& S_{h,s} s_\beta - S_{H,s} c_\beta.
\end{eqnarray}

In terms of the above components, the relevant couplings of the mass eigenstates are  given by:

\begin{eqnarray}
y_{h_i u u}&=&-\frac{m_u }{ \sqrt{2}v s_{\beta }}S_{i,u}\\
y_{h_i d d}&=&-\frac{m_d }{ \sqrt{2} v c_{\beta }}S_{i,d}\\
y_{h_i \chi\chi}&=& \left[\left(g_1 N_{11} - g_2 N_{12} \right) N_{13} +\sqrt{2}\lambda  N_{15} N_{14} \right]
   S_{i,d}\nonumber \\
   & &- \left[\left(g_1 N_{11} - g_2 N_{12} \right) N_{14} - \sqrt{2}\lambda N_{15} N_{13} \right]S_{i,u}   \nonumber \\
   &&+\sqrt{2} 
   \left(\lambda N_{13} N_{14}-\kappa N_{15}^2
  \right)S_{i,s} \, .\nonumber \\
\end{eqnarray}

When $m_A$ is much larger than any of the other mass scales, the singlet components of the non-singlet like Higgs are approximately given by:

\begin{eqnarray}
S_{H,s} &\approx& \frac{\lambda  v  c_{2\beta}}{m_A^2} \left(\frac{2\mu}{s_{2\beta}}+\epsilon \right)\, ,\\
S_{h,s} &\approx& \frac{-\lambda  v \epsilon  s_{2\beta}}{m_{h}^2-m_{h_S}^2},
\end{eqnarray}
where $m_h \sim$ 125 GeV is identified with the SM-like Higgs  and $m_{h_S}$ is the mass of the singlet like Higgs. The up and down components of the singlet-like Higgs are then given as follows:

\begin{eqnarray}
S_{h_S,u} &\approx& \frac{-\lambda  v \epsilon  s_{\beta} s_{2\beta}}{m_{h}^2-m_{h_S}^2}-\frac{\lambda  v  c_\beta c_{2\beta} }{m_A^2} \left(\frac{2\mu}{s_{2\beta}}+\epsilon \right)  \sim   -\frac{\lambda  \mu  v }{m_A^2}\frac{c_{2\beta}}{s_\beta} \\
   \nonumber\\
S_{h_S,d}& \approx&\frac{-\lambda  v \epsilon  c_{\beta} s_{2\beta}}{m_{h}^2-m_{h_S}^2}+\frac{\lambda  v  s_\beta c_{2\beta} }{m_A^2} \left(\frac{2\mu}{s_{2\beta}}+\epsilon \right)  \sim  \frac{\lambda  \mu  v }{m_A^2}\frac{c_{2\beta}}{c_\beta} 
\end{eqnarray}

Decoupling the MSSM-like heavy Higgs, $H$, from the other two, the $2 \times 2$ reduced mass matrix in the  (h,S) basis is given by:
\begin{eqnarray}
\mathcal{M}^2_{hS} & = & \left(
\begin{array}{ccc}
c^2_{2\beta} m_Z^2+\lambda ^2 v^2 s^2_{2\beta}
& 
-\lambda v \epsilon  s_{2\beta}
\\
-\lambda  v \epsilon  s_{2\beta}
& 
\lambda ^2 v^2+\frac{\kappa  \mu}{\lambda }  \left(A_{\kappa }+\frac{4 \kappa  \mu }{\lambda }+3 \mu '\right)-\kappa  \lambda 
   v^2 s_{2\beta}+\frac{\lambda ^2 v^2 \epsilon }{2 \mu } s_{2\beta}+\frac{\lambda}{\mu }  \left(\xi _F \mu '+\xi
   _S\right)
\end{array}
 \right)\nonumber\\
\end{eqnarray}

Generally, since we want to have minimal mixing of the singlet-like Higgs to the others, the (2,2) element of the above should approximately give  the tree-level mass of $h_S$. The (1,1) element should similarly correspond to the mass of the $h$. Loop corrections to this mass should be the usual MSSM corrections from the squarks. Sbottoms and staus generally lead to small ($\sim$ few GeV) negative corrections, while one can get large positive corrections from the stops. Therefore, since we want this $h$  to correspond to the observed SM-like Higgs, we always constrain $\lambda$ to be such that $\mathcal{M}^2_{hS}(1,1) \lesssim 130$ GeV.

In the $Z_3$ NMSSM, $m_3=0$ and therefore  $m_A$ and $A_\lambda$ are not independent parameters, but again related via Eq.~\ref{eq:mAAlambda}. In that case, the (1,3), (2,3) and (3,3) elements of the CP-even mass matrix are as follows: 

\begin{eqnarray}
\mathcal{M}^2_{h_{Z_3}}(1,3)&=&-\lambda  v \mu c_{2\beta} \left(\frac{m_A^2}{2 \mu^2 } s_{2\beta}+\frac{\kappa }{\lambda }\right)\\
\mathcal{M}^2_{h_{Z_3}}(2,3)&=& 2\lambda  v\mu  \left(1-\frac{m_A^2 }{4 \mu^2 }s^2_{2\beta}-\frac{\kappa }{2 \lambda } s_{2\beta}\right)\\
\mathcal{M}^2_{h_{Z_3}}(3,3)&=&\lambda ^2 v^2 s_{2\beta } \left(\frac{m_A^2 s_{2\beta }}{4\mu ^2}-\frac{\kappa }{2\lambda
   }\right)+\frac{\kappa  \mu  A_{\kappa }}{\lambda }+\frac{4 \kappa ^2 \mu ^2}{\lambda ^2} \, .
\end{eqnarray}

When $\kappa/\lambda$ is small,  the CP-even singlet will generally be light and therefore to minimize mixing of the singlet with the SM-like Higgs, we need:

\begin{equation}
m_A^2 =\frac{ 4\mu ^2}{s^2_{2\beta}} \left(1-\frac{\kappa }{2\lambda } s_{2\beta}-\epsilon\right) .
\end{equation}

The above clarifies the limit in which the above reduction is valid: when $2 \mu / s_{2\beta} >> \kappa \mu/\lambda $ the heavy Non-SM like Higgs decouples and with $m_A \sim 2 |\mu| / s_{2\beta}$ the SM-like Higgs has a negligible singlet component.  

The singlet components of the Higgses are now given as: 
\begin{eqnarray}
S_{H,s} &\approx&-\frac{\lambda  v }{2 \mu} c_{2\beta} s_{2\beta}\\
\nonumber \\
S_{h,s} &\approx& \frac{-2\lambda  v \mu \epsilon  }{(m_{h}^2-m_{h_S}^2)}.
\end{eqnarray}

These correspond to the following up and down components of the singlet:

\begin{eqnarray}
S_{h_{S},u} &\approx& \frac{-2\lambda  v \mu \epsilon  }{(m_{h}^2-m_{h_S}^2)} s_\beta +\frac{\lambda  v }{2 \mu} c_{2\beta} s_{2\beta} c_\beta 
\\
S_{h_{S},d} &\approx & \frac{-2\lambda  v \mu \epsilon  }{(m_{h}^2-m_{h_S}^2)} c_\beta -\frac{\lambda  v }{2 \mu} c_{2\beta} s_{2\beta} s_\beta \sim - \frac{\lambda  v }{ 2\mu} c_{2\beta} s_{2\beta} s_\beta
\end{eqnarray}

Note that when $\kappa/\lambda >>1$, $m_{h_S}$ will generically be pushed up and the Singlet Higgs will decouple from the now MSSM like CP-even Higgs sector.

The (2,2) element of the reduced ($2\times 2$) matrix, which in the limit of zero-mixing with the other Higgs should give the tree-level $h_S$ mass, in the $Z_3$ NMSSM is given by:

\begin{eqnarray}
\mathcal{M}^2_{hS_{Z_3}}(2,2)&=&\frac{\kappa  \mu 
  }{\lambda } \left(A_{\kappa }+\frac{4 \kappa  \mu }{\lambda }\right)+\frac{\lambda ^2 v^2 m_A^2}{4 \mu ^2} \left(1-c_{2 \beta }^2\right) s_{2 \beta }^2-\frac{\kappa ^2 \mu ^2 v^2 }{m_A^2}c_{2 \beta
   }^2-\frac{1}{2} \kappa  \lambda  v^2 \left(2 c_{2 \beta }^2+1\right) s_{2 \beta } \nonumber\\
\end{eqnarray}

Setting $\epsilon\sim 0$,  ($m_A = 2 \mu/ s_{2\beta}$):

\begin{eqnarray}
\mathcal{M}^2_{hS_{Z_3}}(2,2)&=&\frac{\kappa  \mu 
  }{\lambda } \left(A_{\kappa }+\frac{4 \kappa  \mu }{\lambda }\right)+\lambda ^2 v^2 \left(1-c_{2 \beta }^2\right)-\frac{\kappa ^2  v^2 }{2} s_{2\beta}^2 c_{2 \beta
   }^2-\frac{1}{2} \kappa  \lambda  v^2 \left(2 c_{2 \beta }^2+1\right) s_{2 \beta } \nonumber\\
\end{eqnarray}

The above in the large $t_\beta$ limit and dropping sub-dominant terms is in agreement with the expressions presented in Refs.~\cite{Draper:2010ew,Carena:2011jy}.

\section{Direct-Detection}
\label{sec:DD}

The spin-independent elastic cross-section for a neutralino scattering off a heavy nucleus due to the exchange of all the Higgses is given by  
\begin{eqnarray}\label{eq:sigSI}
\sigma_{SI} = \frac{4m_r^2}{\pi} \left[Z f_p + (A-Z) f_n\right]^2
\end{eqnarray}
where $m_r = \frac{m_N m_{\chi}}{m_N+m_{\chi}}$, $m_N$ is the mass of the
nucleus, $m_{\chi}$ is the neutralino mass, and in the decoupling limit:
\beq
f_{p,n} = \left(\sum_{q=u,d,s} f_{T_q}^{(p,n)} \frac{a_q}{m_q} + \frac{2}{27}
f_{TG}^{(p,n)} \sum_{q=c,b,t} \frac{a_q}{m_q} \right) m_{(p,n)}, 
\eeq
\begin{eqnarray}
a_u &=&  \frac{-g_2 m_u}{4m_W s_{\beta}} \left[ (g_2 N_{12} - g_1 N_{11}) 
\left\{N_{13} \left[\frac{-S_{h_{S},u} S_{h_{S},d}}{m_{h_S}^2}-s_{\beta} c_{\beta} \left( \frac{1}{m_h^2} - \frac{1}{m_H^2}
\right) \right] + N_{14} \left(\frac{s^2_{\beta}}{m_h^2} +\frac{c^2_{\beta}}{m_H^2} + \frac{S_{h_{S},u}^2}{m_{h_S}^2}
\right)\right\}  \right.  \nonumber\\
&& +\sqrt{2}\lambda 
   \left\{N_{13} N_{14} \left( \frac{- S_{h,s} s_\beta}{m_{h}^2} +\frac{S_{H,s} c_\beta}{m_{H}^2}  +\frac{S_{h_{S},u}S_{h_{S},s}}{m_{h_S}^2}\right) \right. \nonumber \\
 &&  \qquad \qquad  \left. +N_{15} \left[N_{14} \left(  c_\beta s_\beta \left(  \frac{1}{m_h^2}-\frac{1}{m_H^2}\right)  +\frac{S_{h_{S},d} S_{h_{S},u}}{m_{h_S}^2}\right) +N_{13} \left(\frac{s^2_{\beta}}{m_h^2} +\frac{c^2_{\beta}}{m_H^2} + \frac{S_{h_{S},u}^2}{m_{h_S}^2}
\right)
\right]\right\} \nonumber \\
&& \left.    -\sqrt{2} \kappa N_{15}^2 \left( \frac{- S_{h,s} s_\beta}{m_{h}^2} +\frac{S_{H,s} c_\beta}{m_{H}^2}  +\frac{S_{h_{S},u}S_{h_{S},s}}{m_{h_S}^2}\right)  \right] , \nonumber \\
 \label{au:eq} \\
 \nonumber\\
a_d &=&  \frac{g_2 m_d}{4m_W c_{\beta}}  \left[ (g_2 N_{12} -
g_1 N_{11}) 
\left\{ N_{13} \left(\frac{c^2_{\beta}}{m_h^2} +\frac{s^2_{\beta}}{m_H^2} + \frac{S_{h_{S},d}^2}{m_{h_S}^2}
\right) - N_{14}\left[ \frac{S_{h_{S},u} S_{h_{S},d}}{m_{h_S}^2} +c_{\beta} s_{\beta} \left( \frac{1}{m_h^2} - \frac{1}{m_H^2}
\right) \right]\right\} \right. \nonumber \\
&& -\sqrt{2}\lambda 
   \left\{N_{13} N_{14} \left( \frac{- S_{h,s}c_\beta}{m_h^2} -\frac{S_{H,s} s_\beta}{m_H^2} +\frac{S_{h_{S},d}S_{h_{S},s}}{m_{h_S}^2} \right)
   \right. \nonumber \\
  && \left. \qquad \qquad +N_{15} \left[N_{14}\left( \frac{c_\beta^2}{m_h^2} +\frac{s_\beta^2}{m_H^2}+\frac{S_{h_{S},d}^2}{ m_{h_S}^2} \right) +N_{13} \left( c_\beta s_\beta \left( \frac{1}{m_h^2} -\frac{1}{m_H^2} \right) +\frac{S_{h_{S},d} S_{h_{S},u}}{m_{h_S}^2} \right) 
   \right]\right\}  \nonumber \\
&& \left.  +\sqrt{2} \kappa N_{15}^2  \left( \frac{- S_{h,s}c_\beta}{m_h^2} -\frac{S_{H,s} s_\beta}{m_H^2} +\frac{S_{h_{S},d}S_{h_{S},s}}{m_{h_S}^2} \right) \right]\, . \nonumber \\
\label{ad:eq}. 
\end{eqnarray} 
Note that in the non-decoupling limit, the above reproduces the general formula with the replacement of $c_\beta\to -s_\alpha$ and $s_\beta \to c_\alpha$ everywhere except for the common factor.  $m_{(p,n)}$ is either the proton or the neutron mass. For their respective form factors for $\{u, d, s\}$, we use the default parameters used by {\tt{micrOMEGAs\_3.2}}~\cite{Belanger:2013oya,Beringer:1900zz}:
\begin{eqnarray}
f_{T_q}^p = \{0.0153, 0.0191, 0.0447\}; \qquad \qquad f_{T_q}^n =\{0.011, 0.0273, 0.0447\}\;.
\end{eqnarray}
Further, $f_{TG}^{(p,n)} =1-f_{T_u}^{(p,n)}-f_{T_d}^{(p,n)}-f_{T_s}^{(p,n)}$.

\bibliography{GCEBib_revised}
\bibliographystyle{utphys}

\end{document}